\documentclass[%
 reprint,
nofootinbib,
 amsmath,amssymb,
 aps,
]{revtex4-2}

\usepackage{graphicx}
\usepackage{dcolumn}
\usepackage{bm}
\usepackage{braket}
\usepackage{hyperref}
\usepackage[dvipsnames]{xcolor}
\usepackage{slashed}

\usepackage[framemethod=TikZ]{mdframed}
\mdfsetup{roundcorner=5pt}

\newmdenv[
skipabove=5pt,
skipbelow=5pt,
rightline=false,
leftline=false,
topline=false,
bottomline=false,
backgroundcolor=gray!30,
innerleftmargin=5pt,
innerrightmargin=5pt,
innertopmargin=5pt,
innerbottommargin=5pt,
leftmargin=0cm,
rightmargin=0cm,
linewidth=4pt]{greybox}

\newenvironment{slow_eqs}[2][]{%

    \mdfsetup{%
        frametitle={%
            \tikz[baseline=(current bounding box.east),outer sep=0pt]
            \node[anchor=east,rectangle,fill=blue!20]
            {\strut #1};}%
    }%

    \mdfsetup{%
        innertopmargin=0pt,linecolor=blue!20,%
        linewidth=2pt,topline=true,%
        frametitleaboveskip=\dimexpr-\ht\strutbox\relax%
    }
 
\begin{mdframed}[]\relax}{%
\end{mdframed}}

\newcommand{\schro}{Schr\"odinger}

\newcommand{\GeV}{\mathrm{GeV}}

\newcommand{\Mpl}{{\rm M}_{\rm Pl}}

\newcommand{\Hquad}{\hspace{0.4em}} 
\newcommand{\as}{\slashed{a}}

\newcommand{\subheading}[1]{\par\medskip\noindent\textit{#1}\par}

\begin{document}

\preprint{APS/123-QED}

\title{A Tale of Two Gauges: Effective Field Theory \\ for Relativistic Behavior of Cosmological Axions}

\author{Hoang Nhan Luu} \email{hoangnhan.luu@unh.edu}
\author{Chanda Prescod-Weinstein}
\affiliation{Department of Physics \& Astronomy, University of New Hampshire, Durham, NH 03824, USA}

\begin{abstract}

In this work, we present a formalism to model the relativistic behavior of axions, extending the studies of~\cite{Namjoo:2017nia, Salehian:2020bon, Salehian:2021khb}. The relativistic behavior of axions is surprisingly difficult to model precisely, as it involves oscillations on timescales much shorter than the Hubble timescale. To overcome this challenge, one typically resorts to some form of effective treatment, focusing only on the time-averaged description of the exact oscillations. Salehian, Namjoo \& Kaiser~\cite{Salehian:2020bon} provide a systematic framework for such treatment, based on the effective field theory formalism. While the aforementioned study was formulated for axion perturbations in the Newtonian gauge with no anisotropic stress, we extend the formalism to the synchronous gauge that is more conventionally used for numerical implementation in a realistic cosmological setting. Unlike~\cite{Salehian:2020bon}, however, we propose a fluid interpretation in which the axion field can be identified as a perfect fluid at all times, both in the exact and effective regimes. Moreover, we present the effective field theory for the Newtonian gauge with non-zero anisotropic stress, making the original formulation more general and useful for scenarios where the matter content of the universe is multi-component. These results lay the theoretical foundation for a companion paper~\cite{Luu:2026bzr} where we discuss how the axion field should be incorporated alongside other species in common cosmological Boltzmann solvers.

\end{abstract}

\maketitle


\section{Introduction} \label{sec:intro}

Axions and axion-like particles has been proposed as one of the most promising dark matter candidates in the extension of the Standard Model of particle physics~\cite{Peccei:2006as}. In cosmology and astrophysics, axions with an ultralight mass, $m \lesssim 10^{-18}~{\rm eV}$, are of particular interest because of the rich and diverse phenomenologies imprinted on large- and small-scale observations.

On the cosmological scales, axions with $m \sim 10^{-33}~{\rm eV}$ could explain the recently emerging evidence for dark energy with an evolving equation of state (EOS)~\cite{Luu:2025fgw,Wolf:2025jed}, instead of the positive cosmological constant as in the standard $\Lambda$-cold-dark-matter~($\Lambda$CDM) model. On the galactic scales, the hypothetical fuzzy dark matter~(FDM)~\cite{Hu:2000ke,Hui:2021tkt,Eberhardt:2025caq} could be composed of axions with $m \sim 10^{-22}~{\rm eV}$, which helps explain some of the small-scale crisis within the collisionless cold dark matter paradigm~\cite{Bullock:2017xww}. Even in the intermediate mass range of $10^{-30} - 10^{-25}~{\rm eV}$ where the axion field is considered neither dark-matter-like~(DM-like) nor dark-energy-like~(DE-like), its presence at a percent level could explain Lyman-$\alpha$ observations~\cite{Rogers:2023upm} or relieve the Hubble tension by uplifting the electron mass in the early universe when coupled to the Higgs field~\cite{Luu:2021yhl, Baryakhtar:2024rky}.

Ultralight axions are also motivated from a theoretical perspective. In string theory, they appear in the form of pseudoscalar fields that are associated with geometrical properties of the compactified Calabi-Yau manifold~\cite{Svrcek:2006yi,Cicoli:2021gss}. The co-existence of multiple ultralight axion species with non-negligible self-interaction is a generic prediction of string theory~\cite{Arvanitaki:2009fg, Jain:2025vfh}. These axion models have been extensively studied in the literature, and shown to exhibit a variety of interesting implications, {\it e.g.},~\cite{Luu:2018afg, Glennon:2022huu, Glennon:2023gfm, Luu:2024lfq, Luu:2023dmi, Mirasola:2024pmw}.

In general, the equations of motion governing the exact evolution of the cosmological axion field cannot be solved exactly up to the present day ($z = 0$), even with numerical methods, due to its ``Compton'' oscillations, {\it i.e.}, field variations on the natural timescale of its inverse mass. Concretely, let us consider the equation governing the axion background field $\bar{\phi}$
\begin{align}
    \ddot{\bar{\phi}} + 3H\dot{\bar{\phi}} + m^2\bar{\phi} = 0.
\end{align}
It is straightforward to see that this equation describes a damped harmonic oscillator and hence the solution oscillates with frequency $m^{-1}$. That means, unless we are interested in an extremely light field with $m \sim 10^{-33}~{\rm eV}$, the axion field typically oscillates on a much shorter time scale than other dynamical processes that are cosmologically relevant at late times when $t \sim H_0^{-1}$. For instance, fuzzy dark matter of mass $\sim 10^{-22}~{\rm eV}$ has an oscillating period of roughly $m^{-1} \sim 2.5$ months (when converting to SI units). Comparing this period with the Hubble time scale which is on the order of billion years, we would need at least $10^{11}-10^{12}$ cycles to precisely trace these oscillations throughout cosmic history, which is prohibitively expensive in numerical computation. 

For this reason, several studies in the past have been motivated to describe the axion field as an effective fluid, where the Compton oscillations are averaged/smoothed out in some certain ways. For example, some early works, such as \cite{Turner:1983he, Ratra:1990me}, use a simple integral over a period, {\it i.e.},
\begin{align}
    \braket{X(t)} = \dfrac{m}{2\pi}\int_0^{2\pi/m} X(t')dt', \label{eq:cycle_avg}
\end{align}
where $X(t)$ could be any oscillating function such as the axion field itself. As such, it is usually termed ``cycle-averaging'', namely {\it averaging over one cycle}. Importantly, in \cite{Ratra:1990me} and later studies~\cite{Hwang:2009js, Park:2012ru} the authors also made use of the Wentzel–Kramers–Brillouin~(WKB) ansatz by assuming that the axion field could be expressed as
\begin{align}
    \phi(\bm{x},t) = \phi_+(\bm{x},t)\cos(2mt) + \phi_-(\bm{x},t)\sin(2mt) \label{eq:ansatz_phi}
\end{align}
in order to explicitly factor out the rapidly oscillating terms from the slowly-varying fields $\phi_+,\phi_-$. This way, the average operator as defined in \eqref{eq:cycle_avg} only applies to the former, fast components. Furthermore, one can now utilize the ansatz \eqref{eq:ansatz_phi} to derive the approximate equations for the axion field with terms containing different powers of $H/m$, where $H$ denotes the Hubble function. These terms can be ignored at sufficiently late times when $H$ becomes increasingly smaller than $m$, or equivalently, when the axion field enters the \textit{non-relativistic} limit. 

Thanks to its simplicity and ability to provide immediate insights, this strategy with cycle-averaging has been implemented into cosmological codes like \texttt{axionCAMB}~\cite{Hlozek:2014lca} for the canonical axion model or \texttt{AxiCLASS}~\cite{Smith:2019ihp} for the early dark enery model. Both are still widely used among the community nowadays for numerical computations of axion linear perturbations. More recently, Passaglia \& Hu suggested an alternative method based on the similar idea of cycle-averaging for approximating the axion effective fluid~\cite{Passaglia:2022bcr}. However, their solution significantly improves the transition conditions such that the effective variables now reflect the true average values of the exact values, which has always been a persisting issue in the previous approach. Incorporating axions alongside other species, this latter method meaningfully improve the numerical precision when computing important observables of the early universe, such as the matter power spectrum or the angular spectrum of cosmic microwave background (CMB) radiation~\cite{Liu:2024yne}.

Although the idea of cycle-averaging is elegant, applying it in practice requires the axion field to be expressed in terms of simple trigonometric functions as in \eqref{eq:ansatz_phi}. This immediately reveals a limitation of the method when dealing with more complex axion theories in which the field solution is non-linear. In this context, a recent series of studies by Salehian, Namjoo, Kaiser, Guth, Zhang and Amin~\cite{Namjoo:2017nia, Salehian:2020bon, Salehian:2021khb} opted to approach this problem using a field-centric averaging procedure called Effective Field Theory~(EFT) derived from the first principles. Here, the axion field is first re-defined in terms of the ``wavefunction'' $\psi$ with explicit oscillating factors (see Eq.~\eqref{eq:phi} below), similar to the above strategies. Unlike those, however, the wavefunction is not necessarily a slowly-varying function after this step, so the equation of motion derived for $\psi$ is still exactly equivalent to the ones of the original field $\phi$. 

The authors then further expand dynamical variables coupled to the axion field in a series of multiple frequencies and small parameters, as in Eqs.~\eqref{eq:nu_expansion} and \eqref{eq:epsilon_expansion} below. Such expansions only hold in the non-relativistic limit and $H/m$ is also one of these small parameters. The final step of this method employs familiar techniques from perturbation theory to compute higher-order corrections and derive equations for the so-called ``slow modes'', {\it i.e.}, the averaged values of dynamical variables (like the wavefunction) that we are interested in.

Compared to the aforementioned approaches, the EFT for axions offers several advantages. First, in this approach, the equations of motion is inclusive of the backreactions of high-frequency modes, while others are not even formulated to account for these effects. For instance, let us quote the EFT equation governing the metric perturbation $\eta$ as derived in \eqref{eq:slow_psi} below
\begin{multline*}
    \eta'_s = \dfrac{i}{4} \left(\tilde{\psi}^*_s\delta\tilde{\psi}_s - \tilde{\psi}_s\delta\tilde{\psi}^*_s \right) - \dfrac{1}{2}\delta\tilde{U}_{\as,s} \\
    -\dfrac{3}{8}\tilde{H}_s \left(\tilde{\psi}^*_s\delta\tilde{\psi}_s + \tilde{\psi}_s\delta\tilde{\psi}^*_s \right) - \dfrac{ik^2}{16m^2a_s^2} \left(\tilde{\psi}^*_s\delta\tilde{\psi}_s - \tilde{\psi}_s\delta\tilde{\psi}^*_s \right) \\ 
    - \dfrac{1}{16}|\tilde{\psi}_s|^2 h'_s + \mathcal{O}(\epsilon^4),
\end{multline*}
where every term in the second and third row stemming from the backreactions induced by axion oscillations at the next-to-leading order. At first glance, this equation may appear complicated, but it contains no oscillatory components and is therefore trivially solvable using standard numerical methods. 

Second, one can choose to include the higher-order corrections up to any desired order of small parameters to archive results arbitrarily close to the exact equations. Notice how $\mathcal{O}(\epsilon^4)$ in the above equation is used to denote possible higher-order terms. Third, and most crucially, the axion EFT enables the exact reconstruction of cosmological variables with minimal effort, a feature that is not viable in any other approaches. In other words, the complete oscillatory behavior inherent to axion dynamics can be obtained without the need to fully solve stiff differential equations.

Building on the initial study from \cite{Salehian:2020bon}, we continue exploring the axion EFT to its full potential, with the major focus on concrete applications in cosmology. As such, our work is split into two parts. Since the EFT calculations have been demonstrated in the Newtonian gauge~\cite{Salehian:2020bon} before, our emphasis in the current paper is to derive the EFT for axion linear perturbations for the first time in the synchronous gauge. Furthermore, we present our independent computations in the Newtonian gauge, hence confirming the results of \cite{Salehian:2020bon}. 

Although metric perturbations expressed in the synchronous gauge are not as intuitive, they are proven to be well behaved and much more stable when being implemented in common cosmological codes. In fact, the two most popular Boltzmann solvers \texttt{CAMB}~\cite{Howlett:2012mh} and \texttt{CLASS}~\cite{Blas:2011rf} were primarily constructed to work in the synchronous gauge. In that context, the analytical framework developed here will serve as a solid foundation for a novel numerical technique that could reconstruct axion oscillations in a realistic cosmological setting, which is detailed in a companion paper published alongside this one~\cite{Luu:2026bzr}.

The remainder of this paper is organized as follows. In Sec.~\ref{sec:exact}, we review the exact equations describing cosmology with the axion field $\phi$, together with the fluid description treating axions as a perfect fluid with definitive density, pressure and momentum flux. Section \ref{sec:eft_syn} is where the majority of the content lies. We start by introducing the effective field theory applied to the axion field, the concept of slow modes and high-frequency modes in Sec.~\ref{sec:Fourier_expansion} and estimation of small parameters in Sec.~\ref{sec:small_params}. We then follow a step-by-step procedure to compute the equations of motion and fluid quantities associated with these modes in Sec.~\ref{sec:relativistic_corrections} with special attention to the axion wavefunction. We conclude the section by presenting a compilation of key results in Sec.~\ref{sec:complete_eft_syn}. While Sec.~\ref{sec:eft_syn} addresses linear perturbations in the synchronous gauge, Sec.~\ref{sec:eft_New} derives the corresponding results in the Newtonian gauge. We will show how gauge transformations can be utilized (Sec.~\ref{sec:gauge_transform}) to relate the already computed equations/quantities of the slow modes in the synchronous gauge to the ones in this new gauge (Sec.~\ref{sec:complete_eft_New}). Our results are then compared with those obtained by \cite{Salehian:2020bon} at the end. Finally, we wrap up the paper with some helpful discussions about future directions of axion EFT in Sec.~\ref{sec:conclusion}.

In this work, we adopt the convention of Friedmann-Robertson-Walker (FRW) metric perturbations from Ma \& Bertschinger~\cite{Ma:1995ey} for the synchronous gauge, {\it i.e.},
\begin{align}
    &g_{ij} = \delta_{ij} + h_{ij}(\tau,\bm{x}), \\ 
    &h_{ij}(\tau,\bm{k}) = h(\tau,\bm{k}) + 6\eta(\tau,\bm{k}) - 2\eta(\tau,\bm{k})\delta_{ij},
\end{align}
and from Weinberg~\cite{Weinberg:2008zzc} for the Newtonian gauge, {\it i.e.},
\begin{align}
    g_{00} = 1 +2\Phi(\tau,\bm{x}), \quad g_{ii} = 1 - 2\Psi(\tau,\bm{x}),
\end{align}
where other components of the metric tensor $g_{\mu\nu}$ vanish. Perturbations variables will always be written in the momentum space from now on, {\it e.g.}, $h = h(\tau,\bm{k})$. We use $\tau$ to denote the conformal time and $t$ for the proper/cosmic time. Natural units are assumed where $\hbar = c = 1$ and $\Mpl = 1/\sqrt{8\pi G} \sim 2.4 \times 10^{18}~{\rm GeV}$.

\section{Exact equations for axions} \label{sec:exact}

In this section, we review the physics of the cosmological axion field, including its equations of motion at the background and linear perturbative levels. The axion is commonly defined as a pseudoscalar field with a cosine potential given by
\begin{align}
    V(\phi) &= m^2F^2\left[ 1 - \cos\left(\dfrac{\phi}{F}\right) \right] \nonumber \\ 
    &\simeq \dfrac{m^2}{2} \phi^2 - \dfrac{m^2}{24F^2}\phi^4 + \mathcal{O}(\phi^6), \label{eq:axion_potential}
\end{align}
where $\phi$ denotes the axion field, $m$ and $F$ are the axion mass and ``decay constant.'' The latter expansion assumes $\phi < F$ and typically $F \sim 10^{17}~\GeV$ for the axion-like particles with Peccei-Quinn (PQ) symmetry breaking occurring before inflation~\cite{Marsh:2015xka}. For simplicity, we will focus in this work on the case where the {\it self-interacting} term is negligible, {\it i.e.}, $V(\phi) \sim m^2\phi^2/2$ only. The generalization to the self-interaction model should be straightforward.

In an expanding universe, the equation of motion of a scalar field is just the Klein-Gordon equation in the presence of the FRW metric tensor
\begin{align}
    \dfrac{1}{\sqrt{-g}}\partial_\mu(\sqrt{-g} g^{\mu\nu}\partial_\nu)\phi - m^2\phi = 0.
\end{align}
Expanding the axion field $\phi = \bar{\phi} + \delta\phi$, the background field equation is given by
\begin{align}
    \ddot{\bar{\phi}} + 3H\dot{\bar{\phi}} + m^2\bar{\phi} = 0, \label{eq:phi_background}
\end{align}
and the equation governing axion linear perturbations in a general gauge, as shown in \cite{Hu:2003hjx}, is given by
\begin{multline}
    \ddot{\delta\phi} + 3H\delta\dot{\phi} + \left(\dfrac{k^2}{a^2} + m^2\right)\delta\phi \\ - \left( \dot{A} - 3\dot{H}_L - \dfrac{k}{a}B\right)\dot{\bar{\phi}} + 2Am^2\bar{\phi} = 0. \label{eq:phi_perturbative}
\end{multline}
Here, ``dot'' operators denote derivatives with respect to cosmic (proper) time. $A, B$ and $H_L$ are metric perturbations whose values are determined by the chosen gauge. Specifically, we have 
\begin{align}
    A = 0, \quad B = 0, \quad H_L = h/6 \label{eq:gauge_syn}
\end{align}
in the \textit{synchronous gauge} and
\begin{align}
    A = \Phi, \quad B = 0, \quad H_L = -\Psi \label{eq:gauge_new}
\end{align}
in the \textit{Newtonian gauge}.

It is straightforward to see that the axion background and perturbative equations yield oscillatory solutions due to terms proportional to $m^2\bar{\phi}$ or $m^2\delta\phi$. In order to factor out these oscillatory factors, we can define the (classical) axion wavefunction such that
\begin{align}
    &\phi = \dfrac{1}{\sqrt{2m}}(\psi e^{-imt} + \psi^*e^{imt}), \label{eq:phi} \\ 
    &\dot{\phi} = -i\sqrt{\dfrac{m}{2}}(\psi e^{-imt} - \psi^* e^{imt}). \label{eq:phidot}
\end{align}
Here, we have implicitly enforced a constraint, $\dot{\psi}e^{-imt} + \dot{\psi}^* e^{imt} = 0$, in deriving $\dot{\phi}$. This condition can be chosen to reduce redundancy because: (1) $\psi$ is a complex field with two degrees of freedom while $\phi$ is just a real scalar with only one degree of freedom; (2) the definition of $\phi$ is invariant under a $U(1)$ transformation such that $\psi \rightarrow \psi + ie^{imt}\chi$, where $\chi \equiv \dot{\psi}e^{-imt} + \dot{\psi}^* e^{imt}$ can be treated as a global gauge field~\cite{Salehian:2020bon}.

Substituting $\phi$ and $\dot{\phi}$ from \eqref{eq:phi} and \eqref{eq:phidot} into \eqref{eq:phi_background} and \eqref{eq:phi_perturbative}, then expanding $\psi = \bar{\psi} + \delta\psi$, we can re-derive the equations of motion in terms of the wavefunction. First is the background equation
\begin{align}
    \dot{\bar{\psi}} + \dfrac{3}{2}H\bar{\psi} - \dfrac{3}{2}H\bar{\psi}^* e^{2imt} = 0.\label{eq:background_exact}
\end{align}
Then, the perturbation equations are given as:
\subheading{Synchronous gauge:}
\begin{multline}
    \dot{\delta\psi} + \left( \dfrac{3}{2}H + \dfrac{ik^2}{2ma^2} \right)\delta\psi + \dfrac{1}{4}\bar{\psi}\dot{h} \\ - e^{2imt}\left[ \left( \dfrac{3}{2}H - \dfrac{ik^2}{2ma^2} \right)\delta\psi^* + \dfrac{1}{4}\bar{\psi}^*\dot{h} \right] = 0, \label{eq:delta_psi_exact}
\end{multline}
\subheading{Newtonian gauge:}
\begin{multline}
    \dot{\delta\psi} + \left( \dfrac{3}{2}H + \dfrac{ik^2}{2ma^2} \right)\delta\psi + \left(im\Phi - \dfrac{1}{2}\dot{\Phi} - \dfrac{3}{2}\dot{\Psi}  \right)\bar{\psi} \\ - e^{2imt} \left[  \left( \dfrac{3}{2}H - \dfrac{ik^2}{2ma^2} \right) \delta\psi^* \right. \\ \left. - \left(im\Phi + \dfrac{1}{2}\dot{\Phi} + \dfrac{3}{2}\dot{\Psi} \right)\bar{\psi}^* \right] = 0. \label{eq:perturbation_new_exact}
\end{multline}
Although the explicit oscillatory factor $e^{2imt}$ appears in the above equations, these equations for the wavefunction are still exactly equivalent to \eqref{eq:phi_background} and \eqref{eq:phi_perturbative}. Written in this form, however, it becomes clear that the terms containing $e^{2imt}$ are responsible for the intrinsic oscillations in the solution. These terms are normally neglected when deriving the \schro~equation in the non-relativistic limit~\cite{Marsh:2015xka, Hui:2016ltb} for simplicity. When retained, they induce non-trivial backreactions on every dynamical variables, as will be shown in the next section.

In addition to the equations of motion for the scalar field and its perturbations, one must also account for the relationship between spacetime expansion and curvature in the presence of the axion field. Assuming the total matter can be described as a perfect fluid on cosmological scales, we can derive the Friedmann equations governing the background pressure and density
\begin{align}
    &3\Mpl^2 H^2 = \bar{\rho}, \label{eq:Friedmann} \\
    &6\Mpl^2 \left(\dot{H} + H^2\right) = -\bar{\rho} - 3\bar{p}. \label{eq:Friedmann_2}
\end{align}
At the perturbative level, \cite{Ma:1995ey} shows the gauge-dependent forms for the Einstein equations governing the metric perturbations as follows: 
\subheading{Synchronous gauge:}
\begin{align}
    &\Mpl^2\left(\dfrac{2k^2}{a^2}\eta - H\dot{h}\right) = - \delta\rho, \label{eq:hprime_exact} \\
    &2\Mpl^2\dot{\eta} = \dfrac{a}{k^2} (\bar{\rho}+\bar{p})\theta, \label{eq:eta_exact} \\
    &\Mpl^2 \left( \ddot{h} + 3H\dot{h} - \dfrac{2k^2}{a^2}\eta \right) = -3\delta p, \label{eq:delta_p_exact} \\
    &\Mpl^2\left[ \ddot{h} + 6\ddot{\eta} + 3H(\dot{h} + 6\dot{\eta}) - \dfrac{2k^2}{a^2}\eta \right] = -3 (\bar{\rho} + \bar{p})\sigma \label{eq:sigma_exact}.
\end{align}
\subheading{Newtonian gauge:}
\begin{align}
    &2\Mpl^2\left[\dfrac{k^2}{a^2}\Psi + 3H(\dot{\Psi} + H\Phi) \right] = -\delta\rho \label{eq:Poisson_exact_New} \\
    &2\Mpl^2\left(\dot{\Psi} + H\Phi\right) = \dfrac{a}{k^2}(\bar{\rho} + \bar{p})\theta, \label{eq:Psi_exact_New} \\
    &2\Mpl^2\left[ \ddot{\Psi} + 2H(\dot{\Psi} + \dot{\Phi}) \right. \nonumber \\ 
    &\hspace{1cm} \left. + \left( 2\dfrac{\ddot{a}}{a} + H^2 \right)\Phi + \dfrac{k^2}{3a^2}(\Psi - \Phi) \right] = \delta p \label{eq:delta_p_exact_New} \\
    &\Mpl^2 \dfrac{2k^2}{a^2}(\Psi - \Phi) = 3(\bar{\rho} + \bar{p})\sigma. \label{eq:sigma_exact_New}
\end{align}
Here, $\delta\rho$ is the total density perturbation, $\theta$ is the divergence of the total fluid velocity $\theta \equiv ik^ju_j$, and $\sigma$ characterizes the total anisotropic shear of the fluid. For convenience, let us define
\begin{align}
    \delta U \equiv - \dfrac{a}{k^2}(\bar{\rho} + \bar{p})\theta,
\end{align}
and still refer to this variable as ``velocity perturbation'', even though it also includes information about the background quantities.

As a convention, from now on, variables without subscript are used to denote the {\it total} quantities. The subscript ``$a$'' is used to denote {\it axion-related} quantities and ``$\as$'' is used to collectively denote {\it non-axion} species
\begin{gather}
    \bar{\rho} = \bar{\rho}_a + \bar{\rho}_{\as}, \quad \bar{p} = \bar{p}_a + \bar{p}_{\as}, \\
    \delta\rho = \delta\rho_a + \delta\rho_{\as}, \quad \delta p = \delta p_a + \delta p_{\as}, \\
    \delta U = \delta U_a + \delta U_{\as}, \quad (\bar{\rho} + \bar{p})\sigma = (\bar{\rho}_{\as} + \bar{p}_{\as})\sigma_{\as}.
\end{gather}
Note that, as a scalar field, the anisotropic stress of the axion is zero at the linear perturbation level, {\it i.e.}, $\sigma_a = 0$.

In this exact regime, the axion (scalar) field can also be described as a perfect fluid. The associated fluid variables have been derived in previous works, {\it e.g.}, \cite{Hu:2003hjx}, as follows
\begin{align}
    \bar{\rho}_a = \dfrac{1}{2}\dot{\bar{\phi}}^2 + \dfrac{m^2}{2}\bar{\phi}^2, \quad \bar{p}_a = \dfrac{1}{2}\dot{\bar{\phi}}^2 - \dfrac{m^2}{2}\bar{\phi}^2 \label{eq:background_density_pressure}
\end{align}
for the background density and pressure. The perturbative fluid variables are given by
\begin{align}
    &\delta\rho_a = \dot{\bar{\phi}}\dot{\delta\phi} - \dot{\bar{\phi}}^2A + m^2\bar{\phi}\delta\phi, \\
    &\delta p_a = \dot{\bar{\phi}}\dot{\delta\phi} - \dot{\bar{\phi}}^2A - m^2\bar{\phi}\delta\phi, \\
    &\delta U_a = -\dfrac{a}{k}(\bar{\rho}_a + \bar{p}_a)B - \dot{\bar{\phi}}\delta\phi
\end{align}
where $A,B$ are the same as in \eqref{eq:gauge_syn} and \eqref{eq:gauge_new}. Again, we can expand these quantities in terms of the axion wavefunction. Substituting \eqref{eq:phi}, \eqref{eq:phidot} into \eqref{eq:background_density_pressure} we obtain the background density and pressure
\begin{align}
    \bar{\rho}_a = m|\bar{\psi}|^2, \quad \bar{p}_a = -\dfrac{m}{2}(\bar{\psi}^2 e^{-2imt} + \bar{\psi}^{*2} e^{2imt}). \label{eq:background_density_pressure}
\end{align}
Similarly, the perturbative quantities are given by
\subheading{Synchronous gauge:}
\begin{align}
    \delta\rho_a &= m(\bar{\psi}^*\delta\psi + \bar{\psi}\delta\psi^*), \label{eq:perturb_density} \\
    \delta p_a &= -m(\bar{\psi} \delta\psi e^{-2imt} + \bar{\psi}^* \delta\psi^* e^{2imt}). \label{eq:perturb_pressure}
\end{align}
\subheading{Newtonian gauge:}
\begin{align}
    \delta\rho_a &= m(\bar{\psi}^*\delta\psi + \bar{\psi}\delta\psi^*) + \dfrac{m}{2}(\bar{\psi}e^{-imt} - \bar{\psi}^* e^{imt})^2\Phi, \\
    \delta p_a &= -m(\bar{\psi} \delta\psi e^{-2imt} + \bar{\psi}^* \delta\psi^* e^{2imt}) \nonumber \\ 
    &\hspace{2.5cm}  + \dfrac{m}{2}(\bar{\psi}e^{-imt} - \bar{\psi}^* e^{imt})^2\Phi.
\end{align}
Since $B = 0$ in the synchronous and Newtonian gauge, the velocity perturbation takes the same form:
\begin{align}
    \delta U_a &= -\dfrac{i}{2}(\bar{\psi}^* \delta\psi - \bar{\psi}\delta\psi^* - \bar{\psi} \delta\psi e^{-2imt} + \bar{\psi}^* \delta\psi^* e^{2imt}). \label{eq:perturb_velocity}
\end{align}
It is worth noting that although the velocity perturbation looks the same, $\delta\psi$ at a particular spacetime point is not necessarily the same in these two gauges because it is not a gauge-invariant quantity.

Most of the equations presented here, especially the ones in the Newtonian gauge, agree with the previous results in \cite{Salehian:2020bon}. For instance, the authors also derived \eqref{eq:background_exact} and \eqref{eq:perturbation_new_exact}. Our latter equation is slightly different from the one in their work because they have substituted $\dot{\Phi}$ from the Einstein ``00'' equation and $\Psi = \Phi$ was assumed there. We choose to keep $\Psi$ and $\Phi$ as distinct variables so that the equations can be applied in a more general case where the axion field is included with other components such as photons, neutrinos, baryons, and so on.

\section{EFT equations for axions in the synchronous gauge} \label{sec:eft_syn}

This section presents the main findings of our work, namely the effective equations governing the so-called ``slow-mode'' (or time-averaged) variables in the synchronous gauge with appropriate corrections in the non-relativistic limit. 

We are going to focus on the detailed derivation for the effective version of Eq.~\eqref{eq:delta_psi_exact} as an example. The goal is to systematically ``integrate out'' the fast oscillating terms with $e^{2imt}$ in this equation. Since this exact equation admits an oscillating solution with a period of order $m^{-1}$, it is natural to first expand the solution in a Fourier time series with the same period (Sec.~\ref{sec:Fourier_expansion}). Unlike the Fourier expansion in $k$-space, however, the resulting modes of this expansion do not evolve independently. A second expansion is therefore required to find the solution in terms of small parameters'' using perturbation theory (section \ref{sec:small_params}). Within this framework, we then perform several computations in Sec.~\ref{sec:relativistic_corrections} to derive an effective version of Eq.~\eqref{eq:delta_psi_exact}, in which the oscillating terms are replaced by backreactions from high-frequency modes. We conclude with Sec.~\ref{sec:complete_eft_syn}, where the complete set of EFT equations in the synchronous gauge is provided. This final section \ref{sec:complete_eft_syn} is also recommended for readers who wish to reference the final results without delving into the technical details.

\subsection{Mode expansion in time} \label{sec:Fourier_expansion}

The presence of axion oscillations induce backreactions on the evolution of other variables. For instance, at the background level, the Hubble function includes the axion density, $\bar{\rho}_a$. Thus, $H$ also includes oscillating terms that can become significant if the axion dominates the total background density.  Similarly, at the perturbation level, the metric perturbations are partly sourced by the axion perturbative density, $\delta\rho_a$. These variables, {\it i.e.}, $h,\eta$ in the synchronous gauge and $\Phi,\Psi$ in the Newtonian gauge, couple to the baryon density, photon multipoles and similar quantities through their equations of motion. Eventually, all dynamical variables will contain some oscillating contributions from the axion either directly or indirectly.

Therefore, it is reasonable to decompose any time-dependent variable into a Fourier series,
\begin{align}
    X(t) = \sum_{\nu=-\infty}^{\infty} X_\nu (t) e^{i\nu mt}. \label{eq:nu_expansion}
\end{align}
Here $\nu$ is an integer running from $-\infty$ to $\infty$ with $\nu=0$ corresponding to the {\it slow mode}, {\it i.e.}, the mode without a fast-oscillating factor $e^{i\nu mt}$. Following the notation of~\cite{Salehian:2020bon}, we denote this special mode with the subscript ``$s$'', {\it e.g.}, $X_s(t)$. Note that although the higher-frequency modes with $\nu \neq 0$ are oscillatory, their \textit{amplitudes} $X_\nu$ are still ``slowly-varying'' functions, meaning they vary on the Hubble timescale.

We define the {\it time-averaged} value, $\braket{X(t)}$, of a variable $X(t)$ as a weighted integral
\begin{align}
    \braket{X(t)} = \int_{-\infty}^{\infty} X(t')W(t-t')dt'
\end{align}
where the window function $W(t-t')$ is chosen such that
\begin{align}
    W(t) = \dfrac{\sin (mt/2)}{\pi t}. \label{eq:window_function}
\end{align}
With this choice of window function, the time average \textit{coincides} with the slow mode, {\it i.e.}, $\braket{X} = X_s$. More generally, any mode (including the slow mode) with frequency $\nu$ satisfies
\begin{align}
    X_\nu(t) = \braket{X(t)e^{i\nu mt}}. \label{eq:average_operator}
\end{align}
Intuitively, the window function cuts off any contributions with an absolute frequency $|\omega|$ higher than $m/2$, keeping only the slow part of $X_\nu$ in Fourier space where $|\omega| \leq  m/2$ (see \cite{Salehian:2020bon} for more detail). Thus, while it might be confusing at first to see a ``time-averaged'' quantity is still time-dependent, we should think of the bracket operator as a ``smoothing'' mechanism that removes any high-frequency oscillations of the function being applied at a given time, keeping only the slow-moving parts. This is why $X_\nu(t)$ variables are slowly varying and is governed by separate equations from $X_s(t)$.

We will focus on the equation of motion for $\delta\psi$ as an example for how to derive slow-mode equations. Starting from \eqref{eq:delta_psi_exact}, let us re-define the dynamical variables in this equation into their dimensionless counterparts
\begin{align}
\begin{gathered}
    \tilde{\psi} = \dfrac{\psi}{\sqrt{m}\Mpl}, \quad \delta\tilde{\psi} = \dfrac{\delta\psi}{\sqrt{m}\Mpl}, \\ \tilde{H} = \dfrac{H}{m}, \quad r = \dfrac{1}{a^2}, \quad \tilde{t} = mt, \label{eq:tilde_variables}
\end{gathered}
\end{align}
which would transform \eqref{eq:delta_psi_exact} into
\begin{align}
    &\delta\tilde{\psi}' + \left( \dfrac{3}{2}\tilde{H} + \dfrac{ik^2}{2m^2}r \right)\delta\tilde{\psi} + \dfrac{1}{4}\tilde{\psi}h' \nonumber \\ 
    &\hspace{0.5cm} - e^{2i\tilde{t}} \left[ \left( \dfrac{3}{2}\tilde{H} - \dfrac{ik^2}{2m^2 a^2} \right)\delta\tilde{\psi}^* + \dfrac{1}{4}\tilde{\psi}^*h' \right] = 0, \label{eq:delta_psi_exact_rescaled}
\end{align}
where ``prime'' operators denote derivatives with respect to $\tilde{t}$. Next, we will apply the time-averaged operator ``$\braket{}$'' to both sides of \eqref{eq:delta_psi_exact_rescaled} to obtain
\begin{multline}
    \delta\tilde{\psi}_\nu' + i\nu\delta\tilde{\psi}_\nu  \\ 
    + \left( \dfrac{3}{2}\tilde{H}_\alpha + \dfrac{ik^2}{2m^2}r_\alpha \right)\delta\tilde{\psi}_{\nu-\alpha} + \dfrac{1}{4}(h'_\alpha + i\alpha h_\alpha)\tilde{\psi}_{\nu-\alpha} \\
    - \left( \dfrac{3}{2}\tilde{H}_\alpha - \dfrac{ik^2}{2m^2}r_\alpha \right)\delta\tilde{\psi}^*_{2+\alpha-\nu} - \dfrac{1}{4}(h'_\alpha + i\alpha h_\alpha)\tilde{\psi}^*_{2+\alpha-\nu} = 0,\label{eq:delta_phi_exact_nu}
\end{multline}
where any subscript other than $\nu$ is implicitly contracted, {\it e.g.}, $\tilde{H}_\alpha\delta\tilde{\psi}_{\nu-\alpha} \equiv \sum_\alpha \tilde{H}_\alpha\delta\tilde{\psi}_{\nu-\alpha}$. In deriving equations like \eqref{eq:delta_phi_exact_nu} we have made use of some useful mathematical identities as follows
\begin{equation}
    \begin{gathered}
    (\tilde{X}\tilde{Y})_\nu = \tilde{X}_\alpha \tilde{Y}_{\nu -\alpha}, \Hquad ( \tilde{X}e^{i\mu\tilde{t}} )_\nu = \tilde{X}_{\nu-\mu}, \Hquad ( \tilde{X}^* )_\nu = \tilde{X}^*_{-\nu}, \\
    (\tilde{X}')_\nu = \tilde{X}'_\nu + i\nu\tilde{X}, \Hquad (\tilde{X}'')_\nu = \tilde{X}''_\nu + 2i\nu\tilde{X}'_\nu - \nu^2\tilde{X}_\nu,
    \end{gathered}
\end{equation}
which are straightforward to prove from \eqref{eq:nu_expansion} and \eqref{eq:average_operator}.\footnote{Since $X_\nu$ are expected to be slowly-varying, we have $\braket{X_\nu} = X_\nu$ even with $\nu \neq 0$.}

\subsection{Expansion in small parameters} \label{sec:small_params}

Up to now, all derived equations are still exact, {\it i.e.}, no approximation has been assumed. However, the mode equation \eqref{eq:delta_phi_exact_nu} of the perturbative wavefunction cannot be solved in isolation, as it depends on an infinite tower of other modes through terms such as $\delta\tilde{\psi}_{\nu-\alpha}$ or $\delta\tilde{\psi}^*_{2+\alpha-\nu}$. This is where the most crucial approximation must be employed via perturbation theory. 

The main idea of this approximation is to further decompose the high-frequency modes ($\nu \neq 0$) into a series containing terms with ascending orders of magnitude
\begin{align}
   X_\nu \simeq X^{(1)}_\nu +  X^{(2)}_\nu + X^{(3)}_\nu + X^{(4)}_\nu + \cdots. \label{eq:epsilon_expansion}
\end{align}
We have $X^{(n)}_\nu \sim \epsilon^n$, where $\epsilon$ collectively represents some {\it small parameters} (such as those listed in \eqref{eq:slow_mode_order} and \eqref{eq:non-axion_order}). Therefore, higher-order terms in this series become increasingly smaller. Under the assumption that the dynamics of $X^{(n)}_\nu$ are governed solely by terms of the same order $\epsilon^n$, Eq.~\eqref{eq:delta_phi_exact_nu} can be reduced to a set of perturbative equations that are trivial to solve.

Evidently, this assumption is only valid when these ``small'' parameters are indeed small. Identifying what parameters are appropriate to use is therefore a crucial step in the EFT formalism. Some examples are as follows
\begin{equation}
    \begin{gathered}
        \tilde{\psi}_s \sim \mathcal{O}(\epsilon), \quad \tilde{H}_s \sim \mathcal{O}(\epsilon), \quad \dfrac{k^2}{m^2 a_s^2} \sim \mathcal{O}(\epsilon), \\
        \delta\tilde{\psi}_s \sim \mathcal{O}(\epsilon), \quad h'_s \sim \mathcal{O}(\epsilon), \quad \eta_s \sim \mathcal{O}(\epsilon).
    \end{gathered} \label{eq:slow_mode_order}
\end{equation}
High-order terms could be given in terms of a combination of the above parameters, {\it e.g.}, both $\tilde{\psi}^2_s$ and $\tilde{H}_s\tilde{\psi}_s$ are considered to be of order $\mathcal{O}(\epsilon^2)$.

In general, the order of each small parameter is determined by their definition and equations of motion. For instance, it is reasonable to assume that $\phi < \Mpl$ at all times.\footnote{In fact, $\phi < F$ is necessary for the expansion of the potential \eqref{eq:axion_potential} to be valid and $F \sim 10^{17}~{\rm GeV} < \Mpl$.} Since $\phi \sim \psi/\sqrt{m}$ from \eqref{eq:phi}, we can show that $\tilde{\psi} < 1$ from \eqref{eq:tilde_variables}, hence $\tilde{\psi} \sim \mathcal{O}(\epsilon)$. Also, the slow mode is always the dominant contribution in the sum of \eqref{eq:nu_expansion} by definition, which leads to $\tilde{\psi}_s \sim \tilde{\psi} \sim \mathcal{O}(\epsilon)$. Similar arguments can be applied to estimate that $\delta\tilde{\psi}_s \sim \mathcal{O}(\epsilon)$. Additionally, the time derivative of $X_\nu$ is considered one order higher than itself, namely, $X_\nu' \sim \mathcal{O}(\epsilon)X_\nu$, given that these quantities are slowly-varying.

Note that the order in $\epsilon$ reflects the degree of relativistic corrections, which is independent of the hierarchy between different variables. This is why the perturbations and the background field of the wavefunction have the same order of $\epsilon$. For instance, $\tilde{\psi}_s$ may be larger than $\delta\tilde{\psi}_s$ by several orders of magnitude, yet a second-order term in $\epsilon$ such as $\tilde{\psi}_s\delta\tilde{\psi}_s$ is certainly much smaller than either $\tilde{\psi}_s$ or $\delta\tilde{\psi}_s$ individually.

Among those listed in Eq.~\eqref{eq:slow_mode_order}, the values of $\tilde{H}_s$ and $k^2/m^2a_s^2$ may vary significantly over cosmic history. To see this more clearly, let us introduced some new notations 
\begin{align}
\begin{gathered}
    \tilde{\rho} = \dfrac{\bar{\rho}}{m^2\Mpl^2}, \quad \tilde{p} = \dfrac{\bar{p}}{m^2\Mpl^2}, \\
    \delta\tilde{\rho} = \dfrac{\delta\rho}{m^2\Mpl^2}, \quad \delta\tilde{p} = \dfrac{\delta p}{m^2\Mpl^2}, \quad \delta\tilde{U} = \dfrac{\delta U}{m\Mpl^2},
\end{gathered}
\end{align}
and then rewrite Eq.~\eqref{eq:Friedmann} with $\bar{\rho}_a$ given by \eqref{eq:background_density_pressure}
\begin{align}
    &3\tilde{H}^2 = |\tilde{\psi}|^2 + \tilde{\rho}_{\as}.\label{eq:Friedmann_rescaled}
\end{align}
In the early universe, $\tilde{H}$ is predominantly driven by $\tilde{\rho}_{\as}$ which is dominated by the photon density, and only drops below unity as $H \lesssim m$.

Similarly, for a small-scale perturbation mode with large $k$, the variable $k^2/m^2a_s^2$ may not be small upon horizon entry when $a_s$ is still small. If we are interested in a very light axion with small $m$, this issue may even apply for perturbation modes with not-so-large $k$. Consequently, the axion EFT only becomes valid at a much later time in a realistic cosmological setting.

As for $\eta_s$ and $h'_s$, we can simply estimate them with Eqs.~\eqref{eq:hprime_exact} and \eqref{eq:eta_exact}
\begin{align}
    &\tilde{H}h' = \dfrac{2k^2}{m^2a^2}\eta + \tilde{\psi}^*\delta\tilde{\psi} + \tilde{\psi}\delta\tilde{\psi}^* + \delta\tilde{\rho}_{\as}, \label{eq:Einstein_h_rescaled} \\
    &\eta' = \dfrac{i}{4} \left( \tilde{\psi}^* \delta\tilde{\psi} - \tilde{\psi}\delta\tilde{\psi}^* - \tilde{\psi} \delta\tilde{\psi} e^{-2i\tilde{t}} + \tilde{\psi}^* \delta\tilde{\psi}^* e^{2i\tilde{t}} \right) - \dfrac{1}{2}\delta \tilde{U}_{\as}, \label{eq:Einstein_eta_rescaled}
\end{align}
Note that we have rescaled and substituted $\delta\tilde{\rho}_a$ and $\delta \tilde{U}_a$ from \eqref{eq:perturb_density} and \eqref{eq:perturb_velocity} in deriving Eqs.~\eqref{eq:Einstein_h_rescaled} and \eqref{eq:Einstein_eta_rescaled}. From the above equations, we can infer
\begin{align}
    &\tilde{H} h' \sim \tilde{\psi}^*\delta\tilde{\psi} \sim \mathcal{O}(\epsilon^2) \rightarrow h'_s \sim \mathcal{O}(\epsilon), \\ 
    &\eta' \sim \tilde{\psi}\delta\tilde{\psi}^* \sim \mathcal{O}(\epsilon^2) \rightarrow \eta_s \sim \eta \sim \mathcal{O}(\epsilon)
\end{align}
assuming that all terms in these equations are of the same order. For completeness, we list the order-of-magnitude estimation for the ``non-axion'' variables as well
\begin{gather}
    \tilde{\rho}_{\as,s} \sim \mathcal{O}(\epsilon^2), \quad \tilde{p}_{\as,s} \sim \mathcal{O}(\epsilon^2), \\ 
    \delta\tilde{\rho}_{\as,s} \sim \mathcal{O}(\epsilon^2), \quad \delta\tilde{p}_{\as,s} \sim \mathcal{O}(\epsilon^2), \quad \delta\tilde{U}_{\as,s} \sim \mathcal{O}(\epsilon^2). \label{eq:non-axion_order}
\end{gather}
Without invoking detailed equation for each non-axion species, there are no general expressions for these quantities, so we assume them to be of the same order as their axion counterparts.

\subsection{Computation of relativistic corrections} \label{sec:relativistic_corrections}

Let us demonstrate how relativistic corrections for the slow-mode version of Eq.~\eqref{eq:delta_psi_exact} are computed explicitly. Consider the perturbative wavefunction with modes $\nu \neq 0$. Expanding the dynamical variables in \eqref{eq:delta_phi_exact_nu} using \eqref{eq:epsilon_expansion} and keeping only terms of order $\mathcal{O}(\epsilon)$ to obtain
\begin{align}
    i\nu\delta\tilde{\psi}^{(1)}_\nu + \dfrac{i}{4} \alpha h_\alpha^{(0)}\tilde{\psi}_{\nu-\alpha}^{(1)} - \dfrac{i}{4}\alpha h_\alpha^{(0)}\tilde{\psi}^{(1)*}_{2+\alpha-\nu} = 0, \label{eq:delta_psi_1}
\end{align}
where $X^{(n)*}_\nu \equiv \left[ X^{(n)}_\nu\right]^*$. Note that we have assumed $\tilde{H}^{(0)}_\nu, \tilde{\psi}^{(0)}_\nu, \delta\tilde{\psi}^{(0)}_\nu$ to vanish because by definition high-frequency terms ($\nu \neq 0$) are smaller than their slow modes which are of order $\mathcal{O}(\epsilon)$ as in \eqref{eq:slow_mode_order}. Next, we need to know $\tilde{\psi}^{(1)}_\nu$ and $h^{(0)}_\nu$ to solve for $\delta\tilde{\psi}^{(1)}_\nu$. The former can be found from the background equation \eqref{eq:background_exact}, averaging this equation gives
\begin{align}
    \tilde{\psi}_\nu^{'} + i\nu\tilde{\psi}_\nu + \dfrac{3}{2}\tilde{H}_\alpha \tilde{\psi}_{\nu - \alpha} - \dfrac{3}{2}\tilde{H}_\alpha\tilde{\psi}^*_{2+\alpha-\nu} = 0. \label{eq:psi_nu}
\end{align}
It is straightforward to see that $\tilde{\psi}^{(1)}_{\nu} = 0$ for $\nu \neq 0$ as we solve the above equation at the first order in $\mathcal{O}(\epsilon)$. We can also derive the equation for $h_\nu$ from \eqref{eq:Einstein_h_rescaled}
\begin{multline}
    \tilde{H}_\alpha \left[ h'_{\nu-\alpha} + i(\nu-\alpha)h_{\nu-\alpha} \right] \\ = \dfrac{2k^2}{m^2}r_\alpha \eta_{\nu-\alpha} + \tilde{\psi}^*_{-\alpha}\delta\tilde{\psi}_{\nu-\alpha} + \tilde{\psi}_\alpha\delta\tilde{\psi}^*_{\alpha-\nu} + \delta\tilde{\rho}_{\as,\nu}, \label{eq:h_nu}
\end{multline}
which yields
\begin{align}
    i\nu\tilde{H_s}h^{(0)}_\nu + i\sum_{\alpha\neq 0}(\nu-\alpha)\tilde{H}^{(1)}_\alpha h^{(0)}_{\nu-\alpha} = 0,
\end{align}
so $h_\nu^{(0)} = 0$ if $\tilde{H}^{(1)}_\nu = 0$ for $\nu \neq 0$. This is in fact the case because averaging Eq.~\eqref{eq:Friedmann_rescaled} gives
\begin{align}
    3\tilde{H}_\alpha \tilde{H}_{\nu-\alpha} = \tilde{\psi}^*_{-\alpha}\tilde{\psi}_{\nu-\alpha} + \tilde{\rho}_{\as, \nu}. \label{eq:Friedmann_nu}
\end{align}
which implies that the lowest-order equation we can derive is
\begin{align}
    2\tilde{H}_s\tilde{H}^{(1)}_{\nu} + \sum_{\alpha \neq 0,\nu} \tilde{H}^{(1)}_\alpha\tilde{H}^{(1)}_{\nu-\alpha} = 0.
\end{align}
Given $\tilde{H}_s \neq 0$ in general, this relation only holds in case $\tilde{H}^{(1)}_\nu = 0$. Plugging $\tilde{\psi}^{(1)}_{\nu} = 0$ and $h^{(0)} = 0$ back in \eqref{eq:delta_psi_1}, we obtain $\delta\tilde{\psi}^{(1)}_\nu = 0$ for any $\nu \neq 0$.

The first-order result is trivial, let us proceed to the next order. Since we have already known that $\delta\tilde{\psi}^{(1)}_\nu$ and $\tilde{H}^{(1)}_\nu$ vanish, expanding \eqref{eq:delta_phi_exact_nu} at the second order of $\mathcal{O}(\epsilon^2)$ yields
\begin{align}
    i\nu \delta\tilde{\psi}_\nu^{(2)} - \left( \dfrac{3}{2}\tilde{H}_s - \dfrac{ik^2}{2m^2a_s^2} \right)\delta\tilde{\psi}^*_s\delta_{\nu,2} - \dfrac{1}{4}h'_s \tilde{\psi}^*_s\delta_{\nu,2} = 0. \label{eq:delta_psi_second_order}
\end{align}
Here, the Kronecker delta symbol $\delta_{\nu,2}$ appears because there is no non-zero contribution of order $\mathcal{O}(\epsilon^2)$ in the last two terms of \eqref{eq:delta_phi_exact_nu} unless $2 - \nu = 0$, which makes $\delta\tilde{\psi}_{2+\alpha-\nu} \rightarrow \delta\tilde{\psi}_s$ and $\tilde{\psi}_{2+\alpha-\nu} \rightarrow \tilde{\psi}_s$ for $\alpha = 0$. From Eq.~\eqref{eq:delta_psi_second_order} we can easily solve for $\delta\tilde{\psi}^{(2)}_\nu$ as follows
\begin{align}
    \delta\tilde{\psi}_\nu^{(2)} = - \left[ \left( \dfrac{3i}{4}\tilde{H}_s + \dfrac{k^2}{4m^2a_s^2} \right)\delta\tilde{\psi}_s^* + \dfrac{i}{8}h_s'\tilde{\psi}_s^* \right]\delta_{\nu,2}.
\end{align}
Although it is straightforward to keep solving for higher-order corrections in terms of slow-mode variables, this second-order correction is sufficient for our purpose.

Let us consider the case with $\nu = 0$ now, Eq.~\eqref{eq:delta_phi_exact_nu} will become
\begin{multline}
    \delta\tilde{\psi}_s' = - \left( \dfrac{3}{2}\tilde{H}_\alpha + \dfrac{ik^2}{2m^2}r_\alpha \right)\delta\tilde{\psi}_{-\alpha} - \dfrac{1}{4}(h'_\alpha + i\alpha h_\alpha)\tilde{\psi}_{-\alpha} \\ + \left( \dfrac{3}{2}\tilde{H}_\alpha - \dfrac{ik^2}{2m^2}r_\alpha \right)\delta\tilde{\psi}^*_{2+\alpha} + \dfrac{1}{4}(h'_\alpha + i\alpha h_\alpha)\tilde{\psi}^*_{2+\alpha},
\end{multline}
which is simply the equation of motion for $\delta\tilde{\psi}_s$. The right hand side can be expanded up to any desired order of accuracy. As we have known $\delta\tilde{\psi}_\nu$ with $\nu \neq0$ up to $\mathcal{O}(\epsilon^2)$, let us only keep terms in the summations over $\alpha$ that are of order $\mathcal{O}(\epsilon^3)$ or below
\begin{multline}
    \delta\tilde{\psi}_s' = - \left( \dfrac{3}{2}\tilde{H}_s + \dfrac{ik^2}{2m^2a_s^2} \right)\delta\tilde{\psi}_s - \dfrac{1}{4}h'_s\tilde{\psi}_s \\ + \left( \dfrac{3}{2}\tilde{H}_s - \dfrac{ik^2}{2m^2a_s^2} \right)\delta\tilde{\psi}^{(2)*}_2 + \dfrac{3}{2} \left( \tilde{H}^{(2)}_{-2} - \dfrac{ik^2}{2m^2}r^{(1)}_{-2} \right)\delta\tilde{\psi}^*_s \\ + \dfrac{1}{4}h'_s\tilde{\psi}^{(2)*}_2 - \dfrac{i}{2}h^{(2)}_{-2}\tilde{\psi}_s^*. \label{eq:delta_psi_2}
\end{multline}
If only the second-order terms (on the first row) are kept, we can recover the exact Eq.~\eqref{eq:delta_psi_exact} for the slow mode $\delta\tilde{\psi}_s$ in absence of the fast-oscillating terms $e^{2imt}$. Thus, the remaining next-leading order terms (on the second and third row) actually represent the backreactions of the high-frequency mode ($\nu=2$ in this case) on the evolution of the slow mode, as we have mentioned before.

Equation \eqref{eq:delta_psi_2} is not yet solvable because a few variables, {\it i.e.}, $\tilde{\psi}^{(2)}_2, \tilde{H}^{(2)}_{-2}, r^{(1)}_{-2}$ and $h^{(2)}_{-2}$, are still unknown. We can find them from the $\nu \neq 0$ equations again. At $\mathcal{O}(\epsilon^2)$ and $\mathcal{O}(\epsilon^3)$, respectively, \eqref{eq:psi_nu} and \eqref{eq:Friedmann_nu} give
\begin{align}
    &i\nu\tilde{\psi}^{(2)}_\nu - \dfrac{3}{2}\tilde{H_s}\tilde{\psi}^*_s\delta_{\nu,2} = 0, \label{eq:psi_tilde_nu_2} \\
    &6\tilde{H}_s\tilde{H}^{(2)}_\nu = \tilde{\psi}_s^*\tilde{\psi}^{(2)}_\nu + \tilde{\psi}^{(2)*}_{-\nu}\tilde{\psi}_s + \tilde{\rho}^{(3)}_{\as,\nu}. \label{eq:H_tilde_nu_2}
\end{align}
From the first equation above we can easily derive $\tilde{\psi}^{(2)}_\nu$ which will be plugged in the second equation together with $\tilde{\rho}^{(3)}_{\as,\nu}$ to determine $\tilde{H}^{(2)}_\nu$.

As such, we are going to find corrections of the non-axion density via its energy-conservation equation at the background level
\begin{align}
    \dot{\bar{\rho}}_{\as} + 3H(\bar{\rho}_{\as} + \bar{p}_{\as}) = 0,
\end{align}
which, after being rescaled and averaged, yields
\begin{align}
    i\nu\tilde{\rho}_{\as,\nu} = - \tilde{\rho}'_{\as,\nu} -
    3(1 + w_{\as})\tilde{H}_\alpha\tilde{\rho}_{\as,\nu-\alpha}, \label{eq:nu_rho_non_axion}
\end{align}
where we have assumed the non-axion EOS $w_{\as} = \bar{p}_{\as}/\bar{\rho}_{\as}$ is constant\footnote{In the $\Lambda$CDM regime, the EOS is indeed a constant for photon, baryons, CDM and dark energy. Massive neutrinos have an EOS radiation-like at early times and matter-like at late times but the transition is on much longer timescales than axion oscillations and they are also a sub-dominant component at all times.}. Since $\tilde{\rho}_{\as}$ is at least of order $\mathcal{O}(\epsilon^2)$, its lowest-order equations for $\nu \neq 0$ should also be of order $\mathcal{O}(\epsilon^2)$
\begin{align}
    i\nu\tilde{\rho}^{(2)}_{\as,\nu} = 0.
\end{align}
It is trivial to see that $\tilde{\rho}_{\as,\nu}^{(2)} = 0$ and $\tilde{p}_{\as,\nu}^{(2)} = 0$ as well thanks to the relation $\tilde{\rho}^{(n)}_{\as,\nu} = w_a \tilde{\rho}^{(n)}_{\as,\nu}$ at any order $n$. Similarly, the third-order equation for $\tilde{\rho}^{(3)}_\nu$ reads
\begin{align}
    i\nu\tilde{\rho}^{(3)}_{\as,\nu} = 3(1 + w_a)\tilde{H}_s\tilde{\rho}^{(2)}_{\as,\nu},
\end{align}
so the third-order term also vanishes, $\tilde{\rho}^{(3)}_\nu = 0$, for $\nu \neq 0$. With this result, $\tilde{\psi}^{(2)}_{\nu}$ and $\tilde{H}^{(2)}_\nu$ can now be obtained from \eqref{eq:psi_tilde_nu_2} and \eqref{eq:H_tilde_nu_2}
\begin{align}
    &\tilde{\psi}^{(2)}_\nu = -\dfrac{3i}{4}\tilde{H}_s\tilde{\psi}^*_s\delta_{\nu,2}, \\
    &\tilde{H}^{(2)}_\nu = -\dfrac{i}{8}\tilde{\psi}_s^{*2}\delta_{\nu,2} + \dfrac{i}{8} \tilde{\psi}^2_s\delta_{\nu,-2}.
\end{align}
These results provide two of the unknown variables in Eq.~\eqref{eq:delta_psi_2} in terms of slow-mode variables.

We also need to compute corrections for the scale factor $a$ and $a$-dependent quantities because of their explicit appearance in \eqref{eq:delta_psi_2} via the term $k^2/m^2 a^2$. Given $q \equiv a^{-1}, r = a^{-2}$ and $H = \dot{a}/a$, we have the following relations
\begin{align}
    a' &= a\tilde{H} \Hquad \longrightarrow \Hquad a'_\nu + i\nu a_\nu  = \tilde{H}_\alpha a_{\nu-\alpha}, \label{eq:a_nu} \\
    qa' &= \tilde{H} \Hquad \longrightarrow \Hquad q_\alpha[a'_{\nu-\alpha} + i(\nu - \alpha) a_{\nu - \alpha}] = \tilde{H}_\nu, \label{eq:q_nu} \\
    r &= q^2 \Hquad \longrightarrow \Hquad r_\nu = q_\alpha q_{\nu-\alpha}. \label{eq:r_nu}
\end{align}
The idea is that we will solve these equations from the top down: $\tilde{H}_\nu \rightarrow a_\nu \rightarrow q_\nu \rightarrow r_\nu$. In general, the scale factor continuously varies throughout the cosmic history ($a \geq 1$ in the future). While $a_s, q_s$ and $r_s$ should not be treated as ``small'' parameters, their corrections can be. We start with the leading-order corrections
\begin{align}
    &i\nu a^{(1)}_\nu = 0, \quad a^{(1)'}_\nu + i\nu a^{(2)}_\nu = \tilde{H}_s a^{(1)}_\nu + \tilde{H}^{(2)}_\nu a_s, \\
    &q_s a^{(1)'}_\nu + q^{(1)}_\nu a'_s + i\nu q_s a^{(2)}_\nu = \tilde{H}^{(2)}_\nu, \\
    &r^{(1)}_\nu = 2 q^{(1)}_\nu q_s. \label{eq:r_nu_1}
\end{align}
Solving the first two equations, we have
\begin{multline}
    a^{(1)}_\nu = 0 \Hquad \rightarrow \Hquad i\nu a^{(2)}_\nu = \tilde{H}^{(2)}_\nu a_s \\ 
    \rightarrow \Hquad q^{(1)}_\nu a'_s + \tilde{H}^{(2)}_\nu a_s q_s = \tilde{H}^{(2)}_\nu. \label{eq:q_nu_1}
\end{multline}
To find out how $q_s$ is related to $a_s$, we can set $\nu = 0$ in \eqref{eq:q_nu} and expand
\begin{multline}
   q_\alpha a'_{-\alpha} - i \alpha q_\alpha a_{- \alpha} = \tilde{H}_s \\ \rightarrow \Hquad q_s a'_s + \mathcal{O}(\epsilon^2) = a^{-1}_s a'_s \Hquad \rightarrow \Hquad q_s = a^{-1}_s + \mathcal{O}(\epsilon^2).
\end{multline}
Plugging this expression back in \eqref{eq:q_nu_1} and then \eqref{eq:r_nu_1} we find $q^{(1)}_\nu = 0$ and $r^{(1)}_\nu = 0$. Lastly, the slow mode of $r$ can be easily derived as well, setting $\nu = 0$ in \eqref{eq:r_nu} yields
\begin{multline}
    r_s = q_\alpha q_{-\alpha} \Hquad \rightarrow \Hquad r_s = q^2_s + \mathcal{O}(\epsilon^2) \\ 
    \rightarrow \Hquad r_s = a^{-2}_s + \mathcal{O}(\epsilon^2).
\end{multline}
The last expression $r_s \simeq a^{-2}_s$ has been used in several places where $k^2/m^2a_s^2$ appears before, we simply show where its original derivation here.

Finally, let us derive corrections for $h_\nu$ with $\nu \neq 0$. We rescale and average Eq.~\eqref{eq:delta_p_exact} to get
\begin{align}
    &h'' = - 3\tilde{H}h' + \dfrac{2k^2}{m^2a^2}\eta - 3\delta\tilde{p} \label{eq:h_dot_dot} \\
    \rightarrow \Hquad &h''_\nu + 2i\nu h'_\nu - \nu^2h_\nu = -3\tilde{H}_\alpha\left[ h'_{\nu-\alpha} + i(\nu-\alpha)h_{\nu-\alpha} \right] \nonumber \\ 
    &\hspace{1.7cm} + \dfrac{2k^2}{m^2}r_\alpha\eta_{\nu-\alpha} - 3(\delta\tilde{p}_{a,\nu} + \delta\tilde{p}_{\as,\nu}),
\end{align}
where the perturbative axion pressure can be computed from \eqref{eq:perturb_pressure}:
\begin{align}
    \delta\tilde{p}_{a,\nu} =  - \tilde{\psi}_\alpha\delta\tilde{\psi}_{\nu-\alpha+2} - \tilde{\psi}^*_{-\alpha}\delta\tilde{\psi}^*_{\alpha-\nu+2}.
\end{align}
Thus, at the lowest orders, this equation reduces to 
\begin{align}
    &\nu^2 h^{(1)}_\nu = 0, \label{eq:h_nu_1} \\
    &\nu^2 h^{(2)}_\nu = - 3\tilde{\psi}_s\delta\tilde{\psi}_s\delta_{\nu,-2} - 3\tilde{\psi}^*_s\delta\tilde{\psi}^*_s\delta_{\nu,2} + 3\delta \tilde{p}^{(2)}_{\as,\nu}. \label{eq:h_nu_2}
\end{align}
We have substituted $h^{(0)}_\nu = 0$ into the above equations. Here, the remaining unknown is the correction $\delta\tilde{p}^{(2)}_{\as,\nu}$, which can be found via the energy conservation equation for the non-axion density perturbations
\begin{align}
    \delta\tilde{\rho}'_{\as} + 3\tilde{H}(\delta\tilde{\rho}_{\as} + \delta\tilde{p}_{\as}) = \dfrac{k^2}{m^2 a^2} \delta\tilde{U}_{\as} - \dfrac{1}{2}(\tilde{\rho}_{\as} + \tilde{p}_{\as}) h'. \label{eq:non-axion_energy}
\end{align}
Since every term in \eqref{eq:non-axion_energy} is at least of order $\mathcal{O}(\epsilon^3)$, it is obvious that the second-order terms must vanish. More rigorously, we have
\begin{align}
    &\delta\tilde{\rho}'_{\as,\nu} + i\nu\tilde{\rho}_{\as,\nu} = - 3\tilde{H}_\alpha \left( \delta\tilde{\rho}_{\as,\nu-\alpha} + \delta\tilde{p}_{\as,\nu-\alpha} \right) \nonumber \\ 
    & + \dfrac{k^2}{m^2}r_\alpha \delta\tilde{U}_{\as,\nu-\alpha} - \dfrac{1}{2}(\tilde{\rho}_{\as,\alpha} + \tilde{p}_{\as,\alpha}) \left[ h'_{\nu-\alpha} + i(\nu-\alpha)h_{\nu -\alpha} \right].
\end{align}
Since the only possible second-order term on the right-hand side includes $h^{(0)}_\nu$, it implies that $\delta\tilde{\rho}^{(2)}_{\as,\nu}$ must vanish
\begin{align}
    i\nu\delta\tilde{\rho}^{(2)}_{\as,\nu} = -\dfrac{i\nu}{2}(\tilde{\rho}_{\as,s} + \tilde{p}_{\as,s})h^{(0)}_\nu \Hquad \rightarrow \Hquad \delta\tilde{\rho}^{(2)}_{\as,\nu} = 0.
\end{align}
We can then make an argument about the relation between $\delta\rho_{\as}$ and $\delta p_{\as}$ as in the background computation. For \textit{isentropic} perturbations, {\it i.e.}, with conserved entropy, the perturbed pressure and density is related by $\delta p_{\as}/\delta\rho_{\as} = w_{\as} - \dot{w_{\as}}/[3H(1+w_{\as})]$. Given the fact that most of the non-axion species (at least in the $\Lambda$CDM model) has a constant EOS, it is reasonable to assume the corrections of $\delta\tilde{p}_\nu$ is proportional to those of $\delta\tilde{\rho}_\nu$ at every order
\begin{align}
    \delta\tilde{p}^{(2)}_{\as,\nu} \simeq w_{\as} \delta\tilde{\rho}^{(2)}_{\as,\nu} \Hquad \rightarrow \Hquad \delta\tilde{p}^{(2)}_{\as,\nu} = 0. 
\end{align}
Substituting this expression back to \eqref{eq:h_nu_2} we find
\begin{align}
    h^{(2)}_\nu = - \dfrac{3}{4}\left(\tilde{\psi}_s\delta\tilde{\psi}_s\delta_{\nu,-2} + \tilde{\psi}^*_s\delta\tilde{\psi}^*_s\delta_{\nu,2} \right).
\end{align}

With this expression in hand, we have all the necessary ingredients to write a fully solvable equation of motion for the field perturbation with relativistic corrections up to order $\mathcal{O}(\epsilon^4)$. These ingredients, 
\begin{align}
    &\tilde{\psi}^{(2)}_2 = -\dfrac{3i}{4}\tilde{H}_s\tilde{\psi}^*_s, \\ &\delta\tilde{\psi}^{(2)}_2 = -\left( \dfrac{3i}{4}\tilde{H}_s + \dfrac{k^2}{4m^2a_s^2} \right)\delta\tilde{\psi}_s^* - \dfrac{i}{8}h_s'\tilde{\psi}_s^*, \\
    &\tilde{H}^{(2)}_{-2} = \dfrac{i}{8}\tilde{\psi}^2_s, \quad r^{(1)}_{-2} = 0, \quad h^{(2)}_{-2} = -\dfrac{3}{4}\tilde{\psi}_s\delta\tilde{\psi}_s,
\end{align}
allow us to rewrite \eqref{eq:delta_psi_2} as
\begin{multline}
    \delta\tilde{\psi}_s' = - \left( \dfrac{3}{2}\tilde{H}_s + \dfrac{ik^2}{2m^2a_s^2} \right)\delta\tilde{\psi}_s - \dfrac{1}{4}h'_s\tilde{\psi}_s \\ + \left( \dfrac{9i}{8}\tilde{H}^2_s + \dfrac{3i}{8}|\tilde{\psi}_s|^2 + \dfrac{ik^4}{8m^4a_s^4} \right)\delta\tilde{\psi}_s + \dfrac{3i}{16}\tilde{\psi}_s^2\delta\tilde{\psi}^*_s \\ + \left( \dfrac{3i}{8}\tilde{H}_s + \dfrac{k^2}{16m^2 a_s^2} \right)\tilde{\psi}_s h'_s + \mathcal{O}(\epsilon^4). \label{eq:slow_delta_psi}
\end{multline}
This equation is one of the main results of our study. It is obvious that we also need to know the equations of motion governing other slow-mode variables, at both background and perturbation levels, to form a closed and solvable system. For instance, we should at least determine the effective equations governing $\tilde{H}_s,\tilde{\psi}_s$ and $h'_s$ to be able to solve for $\delta\tilde{\psi}_s$. These equations will not be derived in detail because the procedure follows exactly the same steps as above. We only show the final results in next subsection.

\subsection{Complete system of EFT equations} \label{sec:complete_eft_syn}

Up to this point, we have worked through the example of how to calculate effective equations governing slow-mode variables by using the example of $\delta \psi$, and at the end of the previous subsection we showed the final form of this equation in \eqref{eq:slow_delta_psi}. We can now share the complete system of EFT equations in the synchronous gauge.

Let us start by presenting the results for the background variables. Most of these are also found in \cite{Salehian:2020bon}. First, the wavefunction corrections are given by:
\begin{align}
    &\tilde{\psi}^{(1)}_\nu = 0, \quad \tilde{\psi}^{(2)}_\nu = -\dfrac{3i}{4}\tilde{H}_s\tilde{\psi}^*_s\delta_{\nu,2}, \label{eq:psi_corrections_1} \\
    &\tilde{\psi}^{(3)}_\nu = -\dfrac{3}{32}\left[|\tilde{\psi}_s|^2 + 2(\tilde{\rho}_{\as,s} + \tilde{p}_{\as,s}) \right]\tilde{\psi}^*_s\delta_{\nu,2} \nonumber \\
    &\hspace{3.3cm} + \dfrac{3}{32}\tilde{\psi}^3_s\delta_{\nu,-2} - \dfrac{3}{64}\tilde{\psi}^{*3}_s\delta_{\nu,4},
\end{align}
Next, we have the corrections of the Hubble function, the scale factor and its related quantities
\begin{align}
    &\tilde{H}^{(1)}_\nu = 0, \nonumber \\
    &\tilde{H}^{(2)}_\nu = -\dfrac{i}{8}\tilde{\psi}_s^{*2}\delta_{\nu,2} + \dfrac{i}{8} \tilde{\psi}^2_s\delta_{\nu,-2}, \Hquad \tilde{H}^{(3)}_\nu = 0, \label{eq:H_corrections}  \\
    &a^{(1)}_\nu = 0, \Hquad a^{(2)}_\nu = -\dfrac{a_s}{16}\left( \tilde{\psi}^{*2}_s\delta_{\nu,2} + \tilde{\psi}^2_s\delta_{\nu,-2} \right), \label{eq:a_nu_2} \\ 
    &q^{(1)}_\nu = 0, \Hquad q^{(2)}_\nu = -\dfrac{a^{(2)}_\nu}{a_s^2}, \label{eq:q_nu_2} \\
    &r^{(1)}_\nu = 0, \Hquad r^{(2)}_\nu = -2\dfrac{a^{(2)}_\nu}{a_s^3}. \label{eq:r_nu_2}
\end{align}
Finally at the background level, we have the corrections for axion fluid variables given as:
\begin{align}
    &\tilde{\rho}^{(2)}_{a,\nu} = 0, \quad \tilde{\rho}^{(3)}_{a,\nu} = \dfrac{3i}{4}\tilde{H}_s (\tilde{\psi}^2_s\delta_{\nu,-2} - \tilde{\psi}^{*2}_s\delta_{\nu,2}), \\
    &\tilde{\rho}^{(4)}_{a,\nu} = -\dfrac{3}{16}(\tilde{\rho}_{\as,s} + \tilde{p}_{\as,s}) (\tilde{\psi}^2_s\delta_{\nu,-2} + \tilde{\psi}^{*2}_s\delta_{\nu,2}) \nonumber \\ &\hspace{3.5cm} - \dfrac{3}{64}(\tilde{\psi}^4_s\delta_{\nu,-4} + \tilde{\psi}^{*4}_s\delta_{\nu,4}) \\
    &\tilde{p}^{(2)}_{a,\nu} = -\dfrac{1}{2}(\tilde{\psi}^2_s\delta_{\nu,-2} + \tilde{\psi}^{*2}_s\delta_{\nu,2}), \quad \tilde{p}^{(3)}_{a,\nu} = 0, \\
    &\tilde{p}^{(4)}_{a,\nu} = \dfrac{3}{64} ( 6\tilde{H}^2_s + |\tilde{\psi}_s|^2 )(\tilde{\psi}^2_s\delta_{\nu,-2} + \tilde{\psi}^{*2}_s\delta_{\nu,2}) \nonumber \\ &\hspace{3.5cm} - \dfrac{3}{32}(\tilde{\psi}^4_s\delta_{\nu,-4} + \tilde{\psi}^{*4}_s\delta_{\nu,4}),
\end{align}
and similarly for the non-axion fluid variables:
\begin{align}
    &\tilde{\rho}^{(2)}_{\as,\nu} = 0, \quad \tilde{\rho}^{(3)}_{\as,\nu}= 0, \nonumber \\ 
    &\tilde{\rho}^{(4)}_{\as,\nu} = \dfrac{3}{16}(\tilde{\rho}_{\as,s} + \tilde{p}_{\as,s}) (\tilde{\psi}^2_s\delta_{\nu,-2} + \tilde{\psi}^{*2}_s\delta_{\nu,2}), \\
    &\tilde{p}^{(2)}_{\as,\nu} = 0, \quad \tilde{p}^{(3)}_{\as,\nu} = 0, \quad \tilde{p}^{(4)}_{\as,\nu} = w_{\as,\nu}\tilde{\rho}^{(4)}_{\as,\nu}.
\end{align}
Note that we implicitly assume $\nu \neq 0$ whenever a quantity with the generic subscript $\nu$ is shown from now on.

One may wonder as to why the corrections (at the same order) of axion and non-axion quantities look different from each other. This can be explained by the fact that the axion EOS $w_a$ oscillates on the time scale of $m^{-1}$, so it directly induces a backreaction to $\tilde{p}^{(2)}_{a,\nu}$ while $\tilde{p}^{(2)}_{\as,\nu}$ is vanishing at this order. The finite contribution of $\tilde{p}^{(2)}_{a,\nu}$ then propagates to higher-order axion terms, which makes them distinct from non-axion terms as well.

The corrections above allow us to calculate complete slow-mode equations for the background variables. Without going into details the effective equations governing the background wavefunction and the Hubble function to $\mathcal{O}(\epsilon^4)$ are given by:
\begin{widetext}
\begin{slow_eqs}[Background equations]{box:background}
    \begin{align}
        &\tilde{\psi}'_s = - \dfrac{3}{2}\tilde{H}_s\tilde{\psi}_s + \dfrac{3i}{16}\tilde{\psi}_s \left( 3|\tilde{\psi}_s|^2 + 2\tilde{\rho}_{\as,s} \right) - \dfrac{9}{32}\tilde{H}_s\tilde{\psi}_s \left(|\tilde{\psi}_s|^2 + \tilde{\rho}_{\as,s} + \tilde{p}_{\as,s} \right), \label{eq:slow_psi} \\
        &3\tilde{H}^2_s = |\tilde{\psi}_s|^2 + \tilde{\rho}_{\as,s} + \dfrac{3}{32}|\tilde{\psi}_s|^2\left(|\tilde{\psi}_s|^2 + 2\tilde{\rho}_{\as,s} \right). \label{eq:slow_H}
    \end{align}
\end{slow_eqs}
\end{widetext}
We can also write down complete slow-mode background fluid quantities to the same order as follows
\begin{align}
    &\tilde{\rho}_{a,s} = |\tilde{\psi}_s|^2 + \dfrac{3}{16}\left( |\tilde{\psi}_s|^2 + \tilde{\rho}_{\as,s} \right)|\tilde{\psi}_s|^2 + \mathcal{O}(\epsilon^5), \label{eq:slow_rho} \\
    &\tilde{p}_{a,s} = \dfrac{3}{16}|\tilde{\psi}_s|^4 + \dfrac{3}{8}\left( \tilde{\rho}_{\as,s} + \tilde{p}_{\as,s} \right)|\tilde{\psi}_s|^2 + \mathcal{O}(\epsilon^5). \label{eq:slow_p}
\end{align}

Before moving to the linear perturbations, we note that the slow-mode fluid quantities as derived in \eqref{eq:slow_rho} and \eqref{eq:slow_p} are different from the ``effective'' fluid quantities that are derived in \cite{Salehian:2020bon}. Those effective quantities are derived by assuming the slow-mode equations follow ordinary fluid equations. For example, \eqref{eq:slow_psi} can be identified with the energy conservation equation
\begin{align}
    \tilde{\rho}'_{a, {\rm eff}} + 3\tilde{H}_{\rm eff}(\tilde{\rho}_{a, {\rm eff}} + \tilde{p}_{a, {\rm eff}}) = 0,
\end{align}
if the effective density and pressure are given by
\begin{align}
    &\tilde{\rho}_{a,{\rm eff}} = |\tilde{\psi}_s|^2 + \dfrac{3}{32}\left( |\tilde{\psi}_s|^2 + 2\tilde{\rho}_{\as,s} \right)|\tilde{\psi}_s|^2 + \mathcal{O}(\epsilon^5), \\
    &\tilde{p}_{a,{\rm eff}} = \dfrac{9}{32}|\tilde{\psi}_s|^4 + \dfrac{3}{8}\left( \tilde{\rho}_{\as,s} + \tilde{p}_{\as,s} \right)|\tilde{\psi}_s|^2 + \mathcal{O}(\epsilon^5).
\end{align}
On the other hand, expanding the exact energy conservation equation at $\nu = 0$ up to $\mathcal{O}(\epsilon^4)$ we would obtain
\begin{align}
    \tilde{\rho}'_{a,s} + 3\tilde{H}_s(\tilde{\rho}_{a,s} + \tilde{p}_{a,s}) + 3\tilde{H}^{(2)}_2\tilde{p}^{(2)}_{a,-2} + 3\tilde{H}^{(2)}_{-2}\tilde{p}^{(2)}_{a,2} = 0.
\end{align}
We can see how the two equations above are different by terms with $\tilde{p}^{(2)}_{a,\nu} \neq 0$, which explains why $\tilde{\rho}_{a,s}$ and $\tilde{\rho}_{a,{\rm eff}}$ also end up being slightly different from each other. In practice, these differences in the ``effective'' fluid quantities and the slow modes are relatively small because they only come in at $\mathcal{O}(\epsilon^4)$ corrections.

We now present the analytical expressions for linear perturbations, starting with the relativistic corrections for the wavefunction perturbations:
\begin{align}
    &\delta\tilde{\psi}^{(1)}_\nu = 0, \nonumber \\ 
    &\delta\tilde{\psi}_\nu^{(2)} = - \left[ \left( \dfrac{3i}{4}\tilde{H}_s + \dfrac{k^2}{4m^2a_s^2} \right)\delta\tilde{\psi}_s^* + \dfrac{i}{8}h_s'\tilde{\psi}_s^* \right]\delta_{\nu,2}, \label{eq:dpsi_corrections}
\end{align}
and for the metric perturbations:
\begin{align}
    &h^{(0)}_\nu = 0, \quad h^{(1)}_\nu = 0, \nonumber \\ 
    &h^{(2)}_\nu = - \dfrac{3}{4}\left(\tilde{\psi}_s\delta\tilde{\psi}_s\delta_{\nu,-2} + \tilde{\psi}^*_s\delta\tilde{\psi}^*_s\delta_{\nu,2} \right),  \label{eq:hdot_corrections} \\
    &\eta^{(1)}_\nu = 0, \quad \eta^{(2)}_\nu = \dfrac{1}{8} \left( \tilde{\psi}_s\delta\tilde{\psi}_s\delta_{\nu,-2} + \tilde{\psi}^*_s\delta\tilde{\psi}^*_s\delta_{\nu,2} \right). \label{eq:eta_corrections}
\end{align}

Although the detailed computation for $\eta^{(2)}_\nu$ is not shown explicitly here, we will need the condition $\tilde{\sigma}^{(2)}_{\as,\nu} = 0$ satisfied to derive \eqref{eq:eta_corrections}, where $\tilde{\sigma}_{\as} \equiv (\tilde{\rho}_{\as} + \tilde{p}_{\as})\sigma_{\as}$ is the combined and rescaled anisotropic stress of the non-axion species. Unlike other quantities such as $\tilde{p}^{(2)}_{\as,\nu}$ or $\delta\tilde{p}^{(2)}_{\as,\nu}$, there is no obvious reason for $\tilde{\sigma}^{(2)}_{\as,\nu} = 0$ without considering the detailed physics of individual non-axion species. This should therefore be treated as an \textit{ad hoc} assumption for the time being. Without it, many perturbative variables would depend on unknown corrections from non-axion fluid variables. That said, we provide a justification of this relation in App.~\ref{app:non-axion_slow} for photons. The condition $\tilde{\sigma}^{(2)}_{\as,\nu} = 0$ also suggests that
\begin{align}
    h^{(2)}_\nu + 6\eta^{(2)}_\nu = 0,
\end{align}
which can be derived from Eq.~\eqref{eq:sigma_exact} and explains why $\eta^{(2)}_\nu$ takes the form given in \eqref{eq:eta_corrections}. This is an extremely helpful expression to use for gauge transformations of several variables in the next section.

As with the background, the corrections for axion fluid variables at the perturbative level are given as:
\begin{align}
    &\delta\tilde{\rho}^{(2)}_{a,\nu} = 0, \nonumber \\ 
    &\delta\tilde{\rho}^{(3)}_{a,\nu} = \left[ \left( \dfrac{3i}{2}\tilde{H}_s - \dfrac{k^2}{4m^2a_s^2} \right)\tilde{\psi}_s\delta\tilde{\psi}_s + \dfrac{i}{8}\tilde{\psi}^2_sh'_s \right]\delta_{\nu,-2} \nonumber \\
    &\hspace{0.5cm}- \left[ \left( \dfrac{3i}{2}\tilde{H}_s + \dfrac{k^2}{4m^2a_s^2} \right)\tilde{\psi}^*_s\delta\tilde{\psi}^*_s + \dfrac{i}{8}\tilde{\psi}^{*2}_sh'_s \right]\delta_{\nu,2}  \\
    &\delta\tilde{p}^{(2)}_{a,\nu} = -\tilde{\psi}_s\delta\tilde{\psi}_s\delta_{\nu,-2} - \tilde{\psi}^*_s\delta\tilde{\psi}^*_s\delta_{\nu,2}, \quad \delta\tilde{p}^{(3)}_{a,\nu} = 0, \\
    &\delta\tilde{U}^{(2)}_{a,\nu} = \dfrac{i}{2} \left( \tilde{\psi}_s\delta\tilde{\psi}_s\delta_{\nu,-2} - \tilde{\psi}^*_s\delta\tilde{\psi}^*_s\delta_{\nu,2} \right), \nonumber \\
    &\delta\tilde{U}^{(3)}_{a,\nu} = - \left(\dfrac{ik^2}{8m^2a_s^2}\tilde{\psi}_s\delta\tilde{\psi}_s + \dfrac{1}{16}\tilde{\psi}_s^2h'_s \right)\delta_{\nu,-2} \nonumber \\ 
    &\hspace{1.5cm} + \left(\dfrac{ik^2}{8m^2a_s^2}\tilde{\psi}^*_s\delta\tilde{\psi}^*_s - \dfrac{1}{16}\tilde{\psi}^{*2}_sh'_s \right)\delta_{\nu,2},
\end{align}

Meanwhile, the corrections for non-axion fluid variables are vanishing up to the next-leading order
\begin{align}
    &\delta\tilde{\rho}^{(2)}_{\as,\nu} = 0, \quad \delta\tilde{\rho}^{(3)}_{\as,\nu} = - \dfrac{ik^2}{\nu m^2a_s^2}\delta\tilde{U}^{(2)}_{\as,\nu} = 0, \label{eq:delta_rho_nu_2_3} \\
    &\delta\tilde{p}^{(2)}_{\as,\nu} = 0, \quad \delta\tilde{p}^{(3)}_{\as,\nu} \simeq w_{\as} \delta\tilde{\rho}^{(3)}_{\as,\nu} = 0, \label{eq:delta_p_nu_2_3} \\
    &\delta\tilde{U}^{(2)}_{\as,\nu} = -\dfrac{i}{\nu}\tilde{\sigma}^{(2)}_{\as,\nu} = 0, \nonumber \\ 
    &\delta\tilde{U}^{(3)}_{\as,\nu} = -\dfrac{i}{\nu} \left( \tilde{\sigma}^{(3)}_{\as,\nu} - \delta\tilde{p}^{(3)}_{\as,\nu} - \delta\tilde{U}^{(2)'}_\nu - 3\tilde{H}_s\delta\tilde{U}^{(2)}_{\as,\nu}\right) = 0,\label{eq:delta_U_nu_2_3}
\end{align}
where corrections for $\delta\tilde{U}_{\as,\nu}$ can be computed via the momentum conservation equation
\begin{align}
    \delta \tilde{U}'_{\as} + 3\tilde{H}\delta\tilde{U}_{\as} = -\delta\tilde{p}_{\as} + \tilde{\sigma}_{\as}. \label{eq:non-axion_momentum}
\end{align}
Here, we have also shown how higher-order corrections of $\delta\tilde{\rho}_{\as,\nu}$ and $\delta\tilde{U}_{\as,\nu}$ depend on those of $\tilde{\sigma}_{\as,\nu}$ in \eqref{eq:delta_rho_nu_2_3}, \eqref{eq:delta_p_nu_2_3}, \eqref{eq:delta_U_nu_2_3}. It is clear that the former corrections only vanish if the latter ones are assumed to be negligible. Again, we provide an example in App.~\ref{app:non-axion_slow} showing how the corrections to the photon (as a non-axion species) anisotropic stress vanish up to third order, namely $\tilde{\sigma}^{(2)}_\gamma = \tilde{\sigma}^{(3)}_\gamma = 0$.

Repeating the same procedure with $\delta\tilde{\psi}$ and plugging the above corrections from \eqref{eq:psi_corrections_1}, \eqref{eq:H_corrections}, \eqref{eq:dpsi_corrections}, \eqref{eq:hdot_corrections} into Eqs.~\eqref{eq:Einstein_eta_rescaled} and \eqref{eq:Einstein_h_rescaled}, we obtain the slow-mode equations for the metric perturbations
\begin{align}
    &\eta'_s = \dfrac{i}{4} \left(\tilde{\psi}^*_s\delta\tilde{\psi}_s - \tilde{\psi}_s\delta\tilde{\psi}^*_s \right) - \dfrac{1}{2}\delta\tilde{U}_{\as,s} \nonumber \\
    & -\dfrac{3}{8}\tilde{H}_s \left(\tilde{\psi}^*_s\delta\tilde{\psi}_s + \tilde{\psi}_s\delta\tilde{\psi}^*_s \right) - \dfrac{ik^2}{16m^2a_s^2} \left(\tilde{\psi}^*_s\delta\tilde{\psi}_s - \tilde{\psi}_s\delta\tilde{\psi}^*_s \right) \nonumber \\
    &\hspace{3.8cm} - \dfrac{1}{16}|\tilde{\psi}_s|^2 h'_s + \mathcal{O}(\epsilon^4), \label{eq:slow_eta} \\
    &\tilde{H}_s\left( 1 - \dfrac{3}{16}|\tilde{\psi}_s|^2 \right)h'_s = \left[ \dfrac{2k^2}{m^2a_s^2}\eta_s + \tilde{\psi}^*_s\delta\tilde{\psi}_s + \tilde{\psi}_s\delta\tilde{\psi}^*_s \right. \nonumber \\
    &\hspace{0.3cm} \left. + \delta\tilde{\rho}_{\as,s} + \dfrac{3}{16} \left(\tilde{H}^2_s - |\tilde{\psi}_s|^2 \right) \left(\tilde{\psi}^*_s\delta\tilde{\psi}_s + \tilde{\psi}_s\delta\tilde{\psi}^*_s \right) \right. \nonumber \\
    &\hspace{0.8cm} \left. + \dfrac{3ik^2}{16m^2a_s^2}\tilde{H}_s \left( \tilde{\psi}^*_s\delta\tilde{\psi}_s - \tilde{\psi}_s\delta\tilde{\psi}^*_s \right) \right] + \mathcal{O}(\epsilon^4). \label{eq:slow_h}
\end{align}

Lastly, the axion slow-mode fluid variables read
\begin{align}
    &\delta\tilde{\rho}_{a,s} = \tilde{\psi}^*_s\delta\tilde{\psi}_s + \tilde{\psi}_s\delta\tilde{\psi}^*_s + \mathcal{O}(\epsilon^4), \label{eq:slow_delta_rho_a} \\
    &\delta\tilde{p}_{a,s} = \dfrac{k^2}{4m^2a_s^2}\left( \tilde{\psi}^*_s\delta\tilde{\psi}_s + \tilde{\psi}_s\delta\tilde{\psi}^*_s \right) + \mathcal{O}(\epsilon^4), \label{eq:slow_delta_p_a} \\
    &\delta\tilde{U}_{a,s} = -\dfrac{i}{2}\left( \tilde{\psi}^*_s\delta\tilde{\psi}_s - \tilde{\psi}_s\delta\tilde{\psi}^*_s \right) + \dfrac{3}{4}\tilde{H}_s\left( \tilde{\psi}^*_s\delta\tilde{\psi}_s + \tilde{\psi}_s\delta\tilde{\psi}^*_s \right) \nonumber \\ 
    & + \dfrac{ik^2}{8m^2a_s^2}\left( \tilde{\psi}^*_s\delta\tilde{\psi}_s - \tilde{\psi}_s\delta\tilde{\psi}^*_s \right) + \dfrac{1}{8}|\tilde{\psi}_s|^2 h'_s + \mathcal{O}(\epsilon^4). \label{eq:slow_delta_U_a}
\end{align}

Let us also highlight that any dynamical variables can be restored from their slow modes via the two expressions \eqref{eq:nu_expansion} and \eqref{eq:epsilon_expansion}. For instance, the background wavefunction and the Hubble function read
\begin{align}
    \tilde{\psi} &= \tilde{\psi}_s + \left( \tilde{\psi}^{(2)}_2 + \tilde{\psi}^{(3)}_2 \right) e^{2i\tilde{t}} + \tilde{\psi}^{(3)}_{-2} e^{-2i\tilde{t}} + \tilde{\psi}^{(3)}_4 e^{4i\tilde{t}}, \nonumber \\
    &= - \left[ \dfrac{3i}{4}\tilde{H}_s + \dfrac{3}{32}|\tilde{\psi}_s|^2 + \dfrac{3}{16} \left( \tilde{\rho}_{\as,s} + \tilde{p}_{\as,s} \right) \right]\tilde{\psi}^*_s e^{2i\tilde{t}} \nonumber \\ 
    &\hspace{1.5cm} + \dfrac{3}{32}\tilde{\psi}^3e^{-2i\tilde{t}} - \dfrac{3}{64}e^{4i\tilde{t}}\tilde{\psi}_s^{*3} + \mathcal{O}(\epsilon^4), \\
    \tilde{H} &= \tilde{H}_s + \tilde{H}^{(2)}_2 e^{2i\tilde{t}} + \tilde{H}^{(2)}_{-2} e^{-2i\tilde{t}} \nonumber \\
    &= - \dfrac{i}{8} \left( \tilde{\psi}_s^{*2}e^{2i\tilde{t}} - \tilde{\psi}_s^2 e^{-2i\tilde{t}} \right) + \mathcal{O}(\epsilon^4).
\end{align}
The ability to derive such expressions is unique to the EFT formalism, which allows us not only to compute the time-averaged values (slow modes) but also to \textit{reconstruct} the oscillatory features of the corresponding exact variables. Since they will play a crucial role for our numerical implementation in \cite{Luu:2026bzr}, we present similar expressions at the perturbative level as follows
\begin{align}
    &\delta\tilde{\psi} = \delta\tilde{\psi}_s - \left[ \left( \dfrac{3i}{4}\tilde{H}_s + \dfrac{k^2}{4m^2a_s^2} \right)\delta\tilde{\psi}^*_s + \dfrac{i}{8}\tilde{\psi}_s^*h'_s \right]e^{2i\tilde{t}}, \\
    &h' = h'_s - \dfrac{3i}{2} \left( \tilde{\psi}^*_s\delta\tilde{\psi}_s^*e^{2i\tilde{t}} - \tilde{\psi}_s\delta\tilde{\psi}_se^{-2i\tilde{t}} \right), \\
    &\eta = \eta_s + \dfrac{1}{8} \left( \tilde{\psi}^*_s\delta\tilde{\psi}_s^*e^{2i\tilde{t}} + \tilde{\psi}_s\delta\tilde{\psi}_se^{-2i\tilde{t}} \right),
\end{align}
where the relations $(h')_s = h'_s$ and $(h^{'}_\nu)^{(2)} = i\nu h^{(2)}_\nu$ have been utilized.

One might wonder as to why the slow modes of non-axion fluid quantities are not mentioned so far. As seen before, non-axion species only couple indirectly to the axion field via metric variables $h'$ and $\eta'$. This is why their high-frequency corrections are typically vanishing at low orders, {\it e.g.}, see \eqref{eq:delta_rho_nu_2_3}, \eqref{eq:delta_p_nu_2_3}, \eqref{eq:delta_U_nu_2_3}. It is, therefore, reasonable to approximate these slow-mode quantities to be their exact counterparts. To be precise, we have
\begin{align}
\begin{gathered}
    \tilde{\rho}_{\as,s} = \tilde{\rho}_{\as} + \mathcal{O}(\epsilon^4), \quad \tilde{p}_{\as,s} = \tilde{p}_{\as} + \mathcal{O}(\epsilon^4) \\
    \delta\tilde{\rho}_{\as,s} = \delta\tilde{\rho}_{\as} + \mathcal{O}(\epsilon^4), \quad \delta\tilde{p}_{\as,s} = \delta\tilde{p}_{\as} + \mathcal{O}(\epsilon^4), \\ 
    \delta\tilde{U}_{\as,s} = \delta\tilde{U}_{\as} + \mathcal{O}(\epsilon^4), \quad \tilde{\sigma}_{\as,s} = \tilde{\sigma}_{\as} + \mathcal{O}(\epsilon^4)
\end{gathered}
\end{align}
As expected, the slow-mode quantities are very much the same as the exact ones with small corrections of order $\mathcal{O}(\epsilon^4)$. The treatment for non-axion species should be as general as possible so that we do not need to solve for anything additional. Fortunately, that is exactly the case in the standard $\Lambda$CDM model where non-axion dynamics induced by axion oscillations are mostly insignificant. If one would like to extend the EFT formalism to include more exotic species, the first consideration should be whether the dynamics of these new species varies on timescales comparable to that of the axion, {\it i.e.}, $t \sim m^{-1}$.

For ease of reference, we summarize the effective equations for the perturbative slow modes to $\mathcal{O}(\epsilon^3)$ in the synchronous gauge in the box below:
\begin{widetext}
\begin{slow_eqs}[Perturbative equations in the synchronous gauge]{box:synchronous}
\begin{align}
    &\delta\tilde{\psi}_s' = - \left( \dfrac{3}{2}\tilde{H}_s + \dfrac{ik^2}{2m^2a_s^2} \right)\delta\tilde{\psi}_s - \dfrac{1}{4}h'_s\tilde{\psi}_s \nonumber \\ 
    &\hspace{2cm} + \left( \dfrac{9i}{8}\tilde{H}^2_s + \dfrac{3i}{8}|\tilde{\psi}_s|^2 + \dfrac{ik^4}{8m^4a_s^4} \right)\delta\tilde{\psi}_s + \dfrac{3i}{16}\tilde{\psi}_s^2\delta\tilde{\psi}^*_s + \left( \dfrac{3i}{8}\tilde{H}_s + \dfrac{k^2}{16m^2 a_s^2} \right)\tilde{\psi}_s h'_s, \label{eq:slow_delta_psi_box} \\
    &h'_s = \tilde{H}^{-1}_s\left( 1 - \dfrac{3}{16}|\tilde{\psi}_s|^2 \right)^{-1} \left[ \dfrac{2k^2}{m^2a_s^2}\eta_s + \tilde{\psi}^*_s\delta\tilde{\psi}_s + \tilde{\psi}_s\delta\tilde{\psi}^*_s + \delta\tilde{\rho}_{\as,s} \right. \nonumber \\
    &\hspace{2cm} \left. + \dfrac{3}{16} \left(\tilde{H}^2_s - |\tilde{\psi}_s|^2 \right) \left(\tilde{\psi}^*_s\delta\tilde{\psi}_s + \tilde{\psi}_s\delta\tilde{\psi}^*_s \right) + \dfrac{3ik^2}{16m^2a_s^2}\tilde{H}_s \left( \tilde{\psi}^*_s\delta\tilde{\psi}_s - \tilde{\psi}_s\delta\tilde{\psi}^*_s \right) \right], \label{eq:slow_h_box} \\
    &\eta'_s = \dfrac{i}{4} \left(\tilde{\psi}^*_s\delta\tilde{\psi}_s - \tilde{\psi}_s\delta\tilde{\psi}^*_s \right) - \dfrac{1}{2}\delta\tilde{U}_{\as,s} - \dfrac{3}{8}\tilde{H}_s \left(\tilde{\psi}^*_s\delta\tilde{\psi}_s + \tilde{\psi}_s\delta\tilde{\psi}^*_s \right) - \dfrac{ik^2}{16m^2a_s^2} \left(\tilde{\psi}^*_s\delta\tilde{\psi}_s - \tilde{\psi}_s\delta\tilde{\psi}^*_s \right) - \dfrac{1}{16}|\tilde{\psi}_s|^2 h'_s. \label{eq:slow_eta_box}
\end{align}
\end{slow_eqs}
\end{widetext}

Putting the above results together, the background equations \eqref{eq:slow_psi} and \eqref{eq:slow_H}, as well as the perturbative equations \eqref{eq:slow_delta_psi_box}, \eqref{eq:slow_h_box}, \eqref{eq:slow_eta_box} form the complete system of effective equations that are needed in order to numerically solve for dynamical variables $\tilde{\psi}_s, \tilde{H}_s, \delta\tilde{\psi}_s, h'_s, \eta_s$, respectively. More details about the numerical procedure and how to obtain cosmological initial conditions for this system will be discussed in our second paper~\cite{Luu:2026bzr}. As a side note, it is relatively straightforward to derive the slow-mode version of other Einstein equations such as \eqref{eq:delta_p_exact}, \eqref{eq:sigma_exact} but they are usually not required for numerical computations.

\begin{figure*}
    \centering
    \includegraphics[scale=0.67]{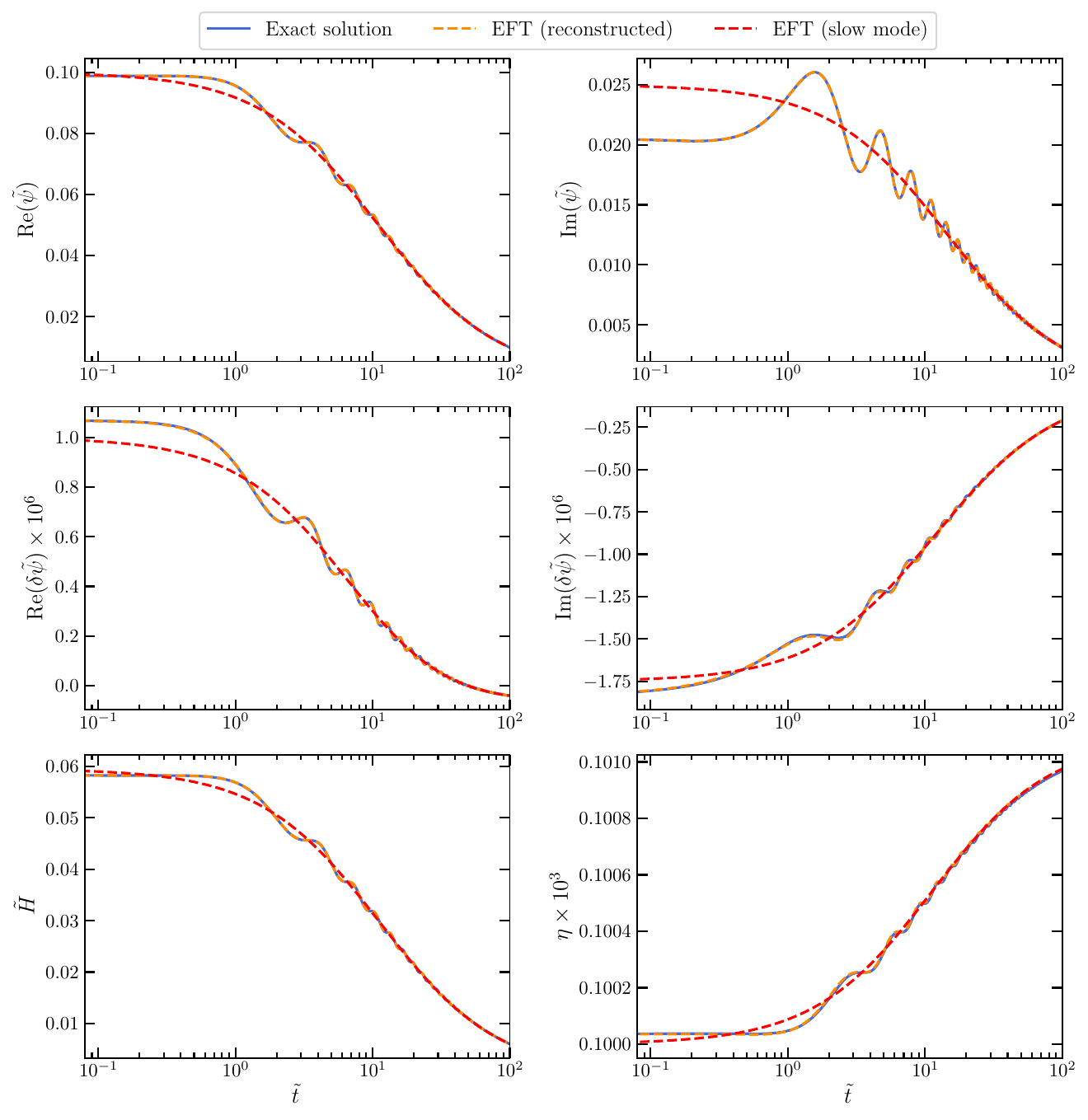} 
    \caption{Reconstructed (orange dashed) and slow-mode (red dashed) variables derived with EFT tested against the exact solution (blue solid). We illustrate the evolution of ${\rm Re}(\tilde{\psi})$ (upper left), ${\rm Im}(\tilde{\psi})$ (upper right), ${\rm Re}(\delta\tilde{\psi})$ (middle left), ${\rm Im}(\delta\tilde{\psi})$ (middle right), $\tilde{H}$ (lower left) and $\eta$ (lower right). Here, perturbations are shown for the wavenumber satisfying $k/m = 0.02$.  The initial conditions of the slow modes are chosen at $\tilde{t}_i = 0$ as follows: $a_{s,i} = 1, \tilde{\psi}_{s,i} = 0.1 + 0.025i, \delta\tilde{\psi}_{s,i} = (1 - 1.75i)\times 10^{-6}, \eta_i = 10^{-4}$. Meanwhile, the initial conditions of the exact solution are computed from the slow-mode ones by incorporating corrections in \eqref{eq:a_nu_2}, \eqref{eq:dpsi_corrections} and \eqref{eq:eta_corrections}. The reconstructed curves are similarly obtained by adding these corrections to the corresponding slow-mode curves.}
    \label{fig:eft_test} 
\end{figure*}

To conclude this section, we show how this solution succeeds by presenting an example illustrating the evolution of several dynamical variables within the EFT framework, compared to their exact counterparts, in Fig.~\ref{fig:eft_test}. The exact solution is obtained by solving Eq.~\eqref{eq:background_exact} for $\tilde{\psi}$ and Eq.~\eqref{eq:Friedmann_rescaled} for $H$ at the background level, as well as Eq.~\eqref{eq:delta_psi_exact} for $\delta\tilde{\psi}$, and Eq.~\eqref{eq:eta_exact} for $\eta$ at the perturbative level. The slow modes are obtained by solving the corresponding effective equations as shown in this section. The initial conditions for these variables are specified in the caption of Fig.~\ref{fig:eft_test}, where they are chosen to be sufficiently small to ensure the validity of the EFT approximation throughout the entire evolution. For simplicity, non-axion species are also not included, so this setup should only be regarded as a toy model. Nonetheless, the results in Fig.~\ref{fig:eft_test} demonstrates that the reconstructed results are in excellent agreement with the exact solution, while the slow modes accurately capture the average behavior of the exact oscillatory variables.

\section{EFT equations for axions in the Newtonian gauge} \label{sec:eft_New}

We began from the position of constructing the EFT in the synchronous gauge because this has not been previously presented in the literature. Instead, Ref.~\cite{Salehian:2020bon} chose to work in the Newtonian gauge when they first came up with the EFT formalism. In that work, \cite{Salehian:2020bon} assume that timelike and spacelike perturbations of the metric are equal, which is to say $\Phi = \Psi$. This is a reasonable assumption, but it only applies to the case where the axion field is the sole component or dominates the total cosmic budget.\footnote{Although a multi-component universe is considered in App.~E in that work, this assumption is still not relaxed.} However, it limits the EFT implementation into a $\Lambda$CDM cosmology, where there will be significant contribution from the anisotropic shear streaming of non-axion species. In this section, we use a gauge transformation to derive the effective equations for slow-mode variables and show that the outcomes agree with those from \cite{Salehian:2020bon} in the limit $\Psi_s \rightarrow \Phi_s$. Therefore, our results derived here for the Newtonian gauge still serve as a new and valuable addition for the EFT formalism.

\subsection{Gauge transformation} \label{sec:gauge_transform}

The first step is to determine the coordinate transformations in terms of metric perturbations. We only need the temporal one which is given by~\cite{Ma:1995ey}
\begin{align}
    \Delta \tau = \dfrac{1}{2k^2} \left(\dfrac{\partial h}{\partial\tau} + 6\dfrac{\partial\eta}{\partial\tau} \right) \Hquad \rightarrow \Hquad \Delta\tilde{t} = \dfrac{m^2 a^2}{2k^2}(h' + 6\eta') \label{eq:Delta_t_gauge} 
\end{align}
where the conformal time has been converted to the proper time as $\Delta\tilde{t} \equiv m\Delta t = ma\Delta\tau$. We then need a relation between metric perturbations in these two gauges
\begin{align}
    &\Psi = \eta -\dfrac{m^2a^2}{2k^2}\tilde{H}(h' + 6\eta'), \label{eq:Psi_gauge} \\
    &\Phi = \dfrac{m^2a^2}{2k^2} \left[ h'' + 6\eta'' +  2\tilde{H}(h' + 6\eta') \right]. \label{eq:Phi_gauge} 
\end{align}
Note that the above equations hold at the same point in 4-dimensional Fourier space $(\tilde{t},\tilde{\bm{k}})$. Lastly, the wavefunction transforms as~\cite{Salehian:2020bon}
\begin{align}
    \Delta\delta\tilde{\psi} = (\tilde{\psi}' - i\tilde{\psi})\Delta\tilde{t} + \dfrac{1}{2}\left( \tilde{\psi} - e^{2i\tilde{t}}\tilde{\psi}^* \right)\Delta\tilde{t}'. \label{eq:psi_gauge}
\end{align}
For convenience, we use the symbols ``tilde'' and ``hat'' to denote the same gauge-dependent perturbative (rescaled) quantities in the synchronous and Newtonian gauges, respectively. For example, the wavefunction perturbations can be written as $\delta\tilde{\psi}$ and $\delta\hat{\psi}$ in these two gauges such that they are related by
\begin{align}
    \delta\hat{\psi}(\tilde{t},\tilde{\bm{k}}) = \delta\tilde{\psi}(\tilde{t},\tilde{\bm{k}}) + \Delta\delta\tilde{\psi}(\tilde{t},\tilde{\bm{k}}),
\end{align}
where $\delta\hat{\psi}(\tilde{t},\tilde{\bm{k}}) \neq \delta\hat{\psi}(\hat{t},\hat{\bm{k}})$ because $\hat{t} = \tilde{t} + \Delta\tilde{t}$ and $\hat{\bm{k}} = \tilde{\bm{k}} + \Delta\tilde{\bm{k}}$. Finally, the gauge transformation implies the following transformations for non-axion fluid quantities~\cite{Weinberg:2008zzc}
\begin{align}
    &\delta\hat{\rho}_{\as} = \delta\tilde{\rho}_{\as} + \tilde{\rho}'_{\as}\Delta\tilde{t}, \nonumber \\ 
    &\delta \hat{p}_{\as} = \delta\tilde{p}_{\as} + \bar{p}'_{\as}\Delta\tilde{t}, \label{eq:nonaxion_gauge} \\ 
    &\delta\hat{U}_{\as} = \delta\tilde{U}_{\as} - (\tilde{\rho}_{\as} + \tilde{p}_{\as})\Delta\tilde{t}. \nonumber
\end{align}

In the next step, we again focus on the equation of motion for the perturbative wavefunction $\delta\hat{\psi}$ with the exact form given by \eqref{eq:perturbation_new_exact} as an example. For the purposes of using gauge transformation, it is not necessary to average this equation and find high-frequency corrections as in the previous section. Instead, we are going to directly convert the slow-mode equation \eqref{eq:slow_delta_psi} already derived in the synchronous gauge. Expanding \eqref{eq:Delta_t_gauge} gives
\begin{align}
    \Delta\tilde{t}_\nu = \dfrac{m^2}{2k^2} b_\alpha \left[ h'_{\nu-\alpha} + \eta'_{\nu-\alpha} + i(\nu - \alpha)(h_{\nu-\alpha} + 6\eta_{\nu-\alpha}) \right],
\end{align}
where a new variable $b$ is defined as
\begin{align}
    b = a^2 \Hquad \rightarrow \Hquad b_\nu = a_\alpha a_{\nu-\alpha}.
\end{align}
Since the lowest non-zero corrections for the scale factor is $a^{(2)}_\nu$ from \eqref{eq:a_nu_2}, we easily find
\begin{align}
    b^{(0)}_\nu = 0, \quad b^{(1)}_\nu = 0 \quad {\rm and} \quad b_s = a^2_s + \mathcal{O}(\epsilon^4),
\end{align}
which implies the relativistic corrections of the temporal transformation given by
\begin{align}
    &\Delta\tilde{t}^{(0)}_\nu = 0, \quad \Delta\tilde{t}^{(1)}_\nu = i\nu\dfrac{m^2a_s^2}{2k^2} \left( h^{(2)}_\nu + 6\eta^{(2)}_\nu \right) = 0, \label{eq:Delta_t_nu} \\
    &\Delta\tilde{t}_s = \dfrac{m^2 a^2_s}{2k^2}(h'_s + 6\eta'_s) + \mathcal{O}(\epsilon^3). \label{eq:Delta_t_s}
\end{align}
Luckily, $\Delta\tilde{t}_s$ has the exact same form as $\Delta\tilde{t}$ in \eqref{eq:Delta_t_gauge} (with $h,\eta$ replaced by $h_s, \eta_s$) up to the second order. This is also the case with $\Psi_s$ and $\Phi_s$, expanding \eqref{eq:Psi_gauge} and \eqref{eq:Phi_gauge} and set $\nu = 0$ gives
\begin{align}
    &\Psi_s = \eta_s - \dfrac{m^2a_s^2}{2k^2} \tilde{H}_s (h'_s + 6\eta'_s) + \mathcal{O}(\epsilon^3), \label{eq:Psi_s_gauge} \\ 
    &\Phi_s = \dfrac{m^2a_s^2}{2k^2} (h''_s + 2\tilde{H}_s h'_s) + \dfrac{3m^2a_s^2}{k^2} (\eta''_s + 2\tilde{H}_s\eta'_s) + \mathcal{O}(\epsilon^3). \label{eq:Phi_s_gauge}
\end{align}
It is easy to see that $\Psi_s \sim \mathcal{O}(\epsilon), \Phi_s \sim \mathcal{O}(\epsilon)$ from the above transformation rules. Similarly, we can compute the slow mode of $\Delta\delta\tilde{\psi}$ from \eqref{eq:psi_gauge}
\begin{align}
    & \Delta\delta\tilde{\psi}_\nu = \left[ \tilde{\psi}'_\alpha + i(\alpha-1)\tilde{\psi}_\alpha \right]\Delta\tilde{t}_{\nu-\alpha} \nonumber \\ 
    &\hspace{0.6cm} + \dfrac{1}{2}\left(\tilde{\psi}_\alpha - \tilde{\psi}^*_{2-\alpha}\right)\left[\Delta\tilde{t}'_{\nu-\alpha} + i(\nu-\alpha)\Delta\tilde{t}_{\nu-\alpha}\right] \\
    &\rightarrow \Hquad\Delta\delta\tilde{\psi}_s = -i\tilde{\psi}_s\Delta\tilde{t}_s + \tilde{\psi}'_s\Delta\tilde{t}_s \nonumber \\ 
    &\hspace{2.5cm} + \dfrac{1}{2}\tilde{\psi}_s\Delta\tilde{t}'_s + i\tilde{\psi}^*_s\Delta\tilde{t}^{(1)}_{-2} + \mathcal{O}(\epsilon^3). \label{eq:Delta_delta_psi_s_0}
\end{align}
There is only one unknown, $\Delta\tilde{t}'_s$, which is computed as
\begin{align}
    \Delta\tilde{t}'_s = \dfrac{m^2a_s^2}{2k^2}(2\tilde{H}_s h'_s + h''_s) = \Phi_s + \mathcal{O}(\epsilon^2), \label{eq:Delta_t_s_prime}
\end{align}
where we have utilized the gauge transformation relation \eqref{eq:Phi_s_gauge} to arrive at the second equality. Substituting $\tilde{\psi}'_s, \Delta\tilde{t}_s, \Delta\tilde{t}'_s, \Delta\tilde{t}^{(1)}_\nu$ from \eqref{eq:slow_psi}, \eqref{eq:Delta_t_s}, \eqref{eq:Delta_t_s_prime}, \eqref{eq:Delta_t_nu}
in Eq.~\eqref{eq:Delta_delta_psi_s_0} yields
\begin{multline}
    \Delta\delta\tilde{\psi}_s = -\dfrac{im^2a_s^2}{2k^2}\tilde{\psi}_s (h'_s + 6\eta'_s) \\ - \dfrac{3m^2a_s^2}{4k^2}\tilde{H}_s\tilde{\psi}_s h'_s + \dfrac{1}{2}\tilde{\psi}_s\Phi_s + \mathcal{O}(\epsilon^3). \label{eq:Delta_delta_psi_s}
\end{multline}
As such, the time derivative of this variable is given by
\begin{multline}
    (\Delta\delta\tilde{\psi}_s)' = \dfrac{3im^2a_s^2}{4k^2}\tilde{H}_s\tilde{\psi}_s( h'_s + 6\eta'_s) - i\tilde{\psi}_s\Phi_s \\ - \dfrac{9}{4}\tilde{H}_s\tilde{\psi}_s\Phi_s + \dfrac{1}{2}\tilde{\psi}_s\Phi'_s
    + \dfrac{9m^2a_s^2}{8k^4}\tilde{H}^2_s\tilde{\psi}_sh'_s - \dfrac{3m^2a_s^2}{4k^2}\tilde{H}'_s\tilde{\psi}_sh'_s \\ + \dfrac{3m^2a_s^2}{16k^2}\left( \dfrac{3}{2}|\tilde{\psi}_s|^2 + \tilde{\rho}_{\as,s} \right)\tilde{\psi}_sh'_s + \mathcal{O}(\epsilon^4).\label{eq:Delta_delta_psi_s_prime}
\end{multline}
Depending on the correction orders of our interest, we should keep track of what terms that need to be included when doing substitution. For example, $\tilde{\psi}'_s = -\frac{3}{2}\tilde{H}_s\tilde{\psi}_s + \mathcal{O}(\epsilon^3)$ should be sufficient for computations of \eqref{eq:Delta_delta_psi_s} but the next-leading-order term is required for \eqref{eq:Delta_delta_psi_s_prime}.

\subsection{Slow-mode equations} \label{sec:complete_eft_New}

Having gauge transformations of all relevant variables, the last step is to compute the equation of motion for $\delta\hat{\psi}_s$. Equation \eqref{eq:slow_delta_psi} can now be written as
\begin{multline}
    \delta\hat{\psi}'_s - (\Delta\delta\tilde{\psi}_s)' = - \left( \dfrac{3}{2}\tilde{H}_s + \dfrac{ik^2}{2m^2a_s^2} \right)(\delta\hat{\psi}_s - \Delta\delta\tilde{\psi}_s) \\ - \dfrac{1}{4}h'_s\tilde{\psi}_s
    + \left( \dfrac{9i}{8}\tilde{H}^2_s + \dfrac{3i}{8}|\tilde{\psi}_s|^2 + \dfrac{ik^4}{8m^4a_s^4} \right)(\delta\hat{\psi}_s - \Delta\delta\tilde{\psi}_s) \\
    + \dfrac{3i}{16}\tilde{\psi}_s^2(\delta\hat{\psi}^*_s - \Delta\delta\tilde{\psi}^*_s) + \left( \dfrac{3i}{8}\tilde{H}_s + \dfrac{k^2}{16m^2 a_s^2} \right)\tilde{\psi}_s h'_s 
\end{multline}
We then substitute $\Delta\delta\tilde{\psi}_s$ and $(\Delta\delta\tilde{\psi}_s)'$ from \eqref{eq:Delta_delta_psi_s} and \eqref{eq:Delta_delta_psi_s_prime} into the above equation. After some straightforward (but lengthy) calculations, we arrive at
\begin{multline}
    \delta\hat{\psi}'_s  = - \left( \dfrac{3}{2}\tilde{H}_s + \dfrac{ik^2}{2m^2a_s^2} \right)\delta\hat{\psi}_s - i\tilde{\psi}_s\Phi_s \\
    + \left( \dfrac{9i}{8}\tilde{H}^2_s + \dfrac{3i}{8}|\tilde{\psi}_s|^2 +  \dfrac{ik^4}{8m^4a_s^4} \right)\delta\hat{\psi}_s + \dfrac{3i}{16}\tilde{\psi}^2_s\delta\hat{\psi}^*_s \\ - \left( \dfrac{3}{2}\tilde{H}_s - \dfrac{ik^2}{4m^2a_s^2} \right)\tilde{\psi}_s\Phi_s + \dfrac{1}{2}\tilde{\psi}_s\Phi'_s + \dfrac{3}{2}\tilde{\psi}_s\eta'_s \\ - \dfrac{3m^2a_s^2}{16k^2}\left( 4\tilde{H}'_s + 3\tilde{H}^2_s - |\tilde{\psi}_s|^2 \right)\tilde{\psi}_sh'_s + \dfrac{3m^2a_s^2}{16k^2}\tilde{\rho}_s\tilde{\psi}_sh'_s. \label{eq:slow_delta_psi_gauge}
\end{multline}
Notice that there is still a mixed combination of metric perturbations from the old and new gauges on the right hand side of this equation. Our goal is to eliminate every terms containing $h_s$ and $\eta_s$, so we need one supplemental equation
\begin{align}
    \Psi'_s = \eta'_s - \tilde{H}_s\Phi_s - \dfrac{m^2a_s^2}{2k^2}\tilde{H}'_s(h'_s + 6\eta'_s) + \mathcal{O}(\epsilon^4), \label{eq:Psi_s_gauge_prime}
\end{align}
which is derived from taking the time derivative of $\Psi_s$ in \eqref{eq:Psi_s_gauge}, followed by a substitution of $\Phi_s$ from \eqref{eq:Phi_s_gauge}. 

Finally, plugging $\eta'_s$ from \eqref{eq:Psi_s_gauge_prime} and $\tilde{H}^2_s$ from \eqref{eq:slow_H} into Eq.~\eqref{eq:slow_delta_psi_gauge} we obtain the slow-mode equation for the perturbative wavefunction in the Newtonian gauge
\begin{multline}
    \delta\hat{\psi}'_s  = - \left( \dfrac{3}{2}\tilde{H}_s + \dfrac{ik^2}{2m^2a_s^2} \right)\delta\hat{\psi}_s - i\tilde{\psi}_s\Phi_s \\
    + \left( \dfrac{3i}{4}|\tilde{\psi}_s|^2 + \dfrac{3i}{8}\tilde{\rho}_{\as,s} + \dfrac{ik^4}{8m^4a_s^4} \right)\delta\hat{\psi}_s + \dfrac{3i}{16}\tilde{\psi}^2_s\delta\hat{\psi}^*_s \\ + \dfrac{ik^2}{4m^2a_s^2} \tilde{\psi}_s\Phi_s + \dfrac{1}{2}\tilde{\psi}_s\Phi'_s + \dfrac{3}{2}\tilde{\psi}_s\Psi'_s + \mathcal{O}(\epsilon^4). \label{eq:slow_delta_psi_New}
\end{multline}
There is a subtle point that is worth paying attention to in this equation. The time derivative on the left hand side is technically taken with respect to $\tilde{t}$, namely $(\Delta\delta\hat{\psi}_s)' = \partial\Delta\delta\hat{\psi}_s/\partial\tilde{t}$. However, the correction to obtain the right time derivative is non-linear, so it can be safely neglected. In other words, we can write $(\Delta\delta\hat{\psi}_s)' \simeq \partial\Delta\delta\hat{\psi}_s/\partial\hat{t}$ and a similar approximation also applies to $\Psi'_s, \Phi'_s$ on the right hand side.

It is straightforward to derive other equations governing the metric perturbations $\Psi_s, \Phi_s$ following a similar procedure. As such, we will show the final results without going into much detail. Let us first compute the slow modes of some non-axion fluid variables in the new gauge from \eqref{eq:nonaxion_gauge}
\begin{align}
    &\delta\hat{\rho}_{\as,s} = \delta\tilde{\rho}_{\as,s} + \tilde{\rho}'_{\as,s}\Delta\tilde{t}_s \nonumber \\
    &\hspace{0.5cm} = \delta\tilde{\rho}_{\as,s} - \dfrac{3m^2a_s^2}{2k^2}\tilde{H}_s(\tilde{\rho}_{\as,s} + \tilde{p}_{\as,s})h'_s + \mathcal{O}(\epsilon^4), \\
    &\delta\hat{U}_{\as,s} = \delta\tilde{U}_{\as,s} - (\tilde{\rho}_{\as,s} + \tilde{p}_{\as,s})\Delta\tilde{t}_s \nonumber \\
    &= \delta\tilde{U}_{\as,s} - \dfrac{m^2a_s^2}{2k^2}(\tilde{\rho}_{\as,s} +\tilde{p}_{\as,s})(h'_s+6\eta'_s) + \mathcal{O}(\epsilon^4).
\end{align}
In order to obtain the equation of motion for $\Psi_s$, we substitute $\eta'_s$ from Eq.~\eqref{eq:slow_eta} into Eq.~\eqref{eq:Psi_s_gauge_prime} and make a transformation for $\delta\tilde{\psi}_s \rightarrow \delta\hat{\psi}_s$ as well as $\delta\tilde{U}_{\as,s} \rightarrow \delta\hat{U}_{\as,s}$
\begin{multline}
    \Psi'_s = -\tilde{H}_s\Phi_s + \dfrac{i}{4} \left(\tilde{\psi}^*_s\delta\hat{\psi}_s - \tilde{\psi}_s\delta\hat{\psi}^*_s \right) \\ - \dfrac{1}{2}\delta\hat{U}_{\as,s} - \dfrac{3}{8}\tilde{H}_s \left(\tilde{\psi}^*_s\delta\hat{\psi}_s + \tilde{\psi}_s\delta\hat{\psi}^*_s \right) \\ - \dfrac{ik^2}{16m^2a_s^2} \left(\tilde{\psi}^*_s\delta\hat{\psi}_s - \tilde{\psi}_s\delta\hat{\psi}^*_s\right) + \mathcal{O}(\epsilon^4). \label{eq:slow_Psi}
\end{multline}
Note that we have also made use of a relation $\tilde{H}'_s = - \frac{1}{2} ( |\tilde{\psi}_s|^2 + \tilde{\rho}_s + \tilde{p}_s ) + \mathcal{O}(\epsilon^4)$ computed from \eqref{eq:slow_H} to eliminate $\tilde{H}'_s$ in deriving the above equation.

Next, we will compute the effective Poisson equation with relativistic corrections. Multiplying both sides of \eqref{eq:Psi_s_gauge} to a factor of $k^2/m^2a_s^2$, then substitute $\eta'_s, \tilde{H}_sh'_s$ from Eqs.~\eqref{eq:slow_eta},\eqref{eq:slow_h} (leading-order terms only) and make a transformation for $\delta\tilde{\psi}_s \rightarrow \delta\hat{\psi}_s$ as well as $\delta\tilde{\rho}_{\as,s} \rightarrow \delta\hat{\rho}_{\as,s}$ to obtain
\begin{multline}
    \dfrac{k^2}{m^2a_s^2}\Psi_s = - \dfrac{1}{2}\left(\tilde{\psi}^*_s\delta\hat{\psi}_s + \tilde{\psi}_s\delta\hat{\psi}^*_s + \delta\hat{\rho}_{\as,s} \right) \\
    -\dfrac{3i}{4}\tilde{H}_s \left(\tilde{\psi}^*_s\delta\hat{\psi}_s - \tilde{\psi}_s\delta\hat{\psi}^*_s\right) + \dfrac{1}{2}|\tilde{\psi}_s|^2\Phi_s \\ + \dfrac{3}{2}\tilde{H}_s\delta\hat{U}_{\as,s} + \mathcal{O}(\epsilon^4). \label{eq:slow_Poisson}
\end{multline}

Lastly, we present the slow modes of the axion fluid variables in the Newtonian gauge
\begin{align}
    &\delta\hat{\rho}_{a,s} = \tilde{\psi}^*_s\delta\hat{\psi}_s + \tilde{\psi}_s\delta\hat{\psi}^*_s - |\tilde{\psi}_s|^2\Phi_s + \mathcal{O}(\epsilon^4), \\
    &\delta\hat{p}_{a,s} = \dfrac{k^2}{4m^2a_s^2}\left( \tilde{\psi}^*_s\delta\hat{\psi}_s + \tilde{\psi}_s\delta\hat{\psi}^*_s \right) + \mathcal{O}(\epsilon^4), \\
    &\delta\hat{U}_{a,s} = -\dfrac{i}{2}\left( \tilde{\psi}^*_s\delta\hat{\psi}_s - \tilde{\psi}_s\delta\hat{\psi}^*_s \right) + \dfrac{3}{4}\tilde{H}_s\left( \tilde{\psi}^*_s\delta\hat{\psi}_s + \tilde{\psi}_s\delta\hat{\psi}^*_s \right) \nonumber \\ 
    &\hspace{1.5cm} + \dfrac{ik^2}{8m^2a_s^2}\left( \tilde{\psi}^*_s\delta\hat{\psi}_s - \tilde{\psi}_s\delta\hat{\psi}^*_s \right) + \mathcal{O}(\epsilon^4),
\end{align}
which are computed by applying gauge transformations to both sides of equations \eqref{eq:slow_delta_rho_a}, \eqref{eq:slow_delta_p_a} and \eqref{eq:slow_delta_U_a}. Up to the next-leading order, our fluid description appears to be consistent with that of~\cite{Salehian:2020bon}, as the slow-mode quantities $\delta\hat{\rho}_{a,s}$, $\delta\hat{p}_{a,s}$, $\delta\hat{U}_{a,s}$ agree with those in their effective fluid description. This is interesting but not surprising because even at the background level, the discrepancy only emerges at the next-to-next-leading corrections for $\tilde{\rho}_{a,s}$ and $\tilde{p}_{a,s}$, as noted in Sec.~\ref{sec:complete_eft_syn}.

The above results are not the only ones that we find agreement with Ref.~\cite{Salehian:2020bon}. Assuming $\Psi_s = \Phi_s$, Eqs.~\eqref{eq:slow_Psi} and \eqref{eq:slow_Poisson} perfectly match Eqs.~(E.12) and (E.13) in the $k$-space of \cite{Salehian:2020bon}. Furthermore, under this assumption, we can take $\Psi'_s = \Phi'_s$ from \eqref{eq:slow_Psi} as well as $(k^2/m^2a_s^2)\Psi_s = (k^2/m^2a_s^2)\Phi_s$ from \eqref{eq:slow_Poisson} and substitute them into the wavefunction equation \eqref{eq:slow_delta_psi_New} to obtain
\begin{multline}
    \delta\hat{\psi}'_s  = - \left( \dfrac{3}{2}\tilde{H}_s + \dfrac{ik^2}{2m^2a_s^2} \right)\delta\hat{\psi}_s - i\tilde{\psi}_s\Phi_s \\
    \qquad + \left( \dfrac{9i}{8}|\tilde{\psi}_s|^2 + \dfrac{3i}{8}\tilde{\rho}_s + \dfrac{ik^4}{8m^4a_s^4} \right)\delta\hat{\psi}_s - \dfrac{7i}{16}\tilde{\psi}^2_s\delta\hat{\psi}^*_s \\ - \left( \delta\hat{U}_s + \dfrac{i}{8}\delta\hat{\rho}_s + 2\tilde{H}_s\Phi_s \right)\tilde{\psi}_s + \mathcal{O}(\epsilon^4), \label{eq:slow_delta_psi_New_2}
\end{multline}
which exactly reproduces Eq.~(E.11) for a multicomponent universe in \cite{Salehian:2020bon}.

The remaining question is whether the difference between these two metric perturbations in the Newtonian gauge can be truly ignored. To gain a better understanding, let us rescale and have a look at the slow mode of Eq.~\eqref{eq:sigma_exact_New}
\begin{align}
    2(\Psi - \Phi) = 3\tilde{\sigma} \Hquad \rightarrow \Hquad \Psi_s - \Phi_s = \dfrac{3}{2}\tilde{\sigma}_{\as,s}.
\end{align}
Here, we should emphasize that although $\tilde{\sigma}^{(2)}_{\as,\nu} = 0$ has been assumed previously, it does not imply the slow-mode quantity like $\tilde{\sigma}_{\as,s}$ also negligible. In fact, since $\tilde{\sigma}_{\as} \equiv (\tilde{\rho}_{\as} + \tilde{p}_{\as})\sigma_{\as}$, it is reasonable to assume that at least $\tilde{\sigma}_{\as,s} \sim \mathcal{O}(\epsilon^2)$ in case $\sigma_{\as} \sim \mathcal{O}(1)$. As we repeat the above calculations, additional terms including $\tilde{\sigma}_{\as,s}$ would enter the relations $\Psi'_s \rightarrow \Phi'_s$ and $(k^2/m^2a_s^2)\Psi_s \rightarrow (k^2/m^2a_s^2)\Phi_s$ at order $\mathcal{O}(\epsilon^3)$. Consequently, these changes only modify Eq.~\eqref{eq:slow_delta_psi_New_2} at order $\mathcal{O}(\epsilon^4)$, which make it valid even when finite anisotropic stress of non-axion species is present.

To recap, let us revert the notation for the perturbative rescaled variables from ``tilde'' back to ``hat'', {\it e.g.}, $\delta\hat{\psi}_s \rightarrow \delta\tilde{\psi}_s$. The system of effective equations for the perturbative slow modes to $\mathcal{O}(\epsilon^3)$ in the Newtonian gauge is then given by:
\begin{widetext}
\begin{slow_eqs}[Perturbative equations in the Newtonian gauge]{box:newtonian}
\begin{align}
    &\delta\tilde{\psi}'_s  = - \left( \dfrac{3}{2}\tilde{H}_s + \dfrac{ik^2}{2m^2a_s^2} \right)\delta\tilde{\psi}_s - i\tilde{\psi}_s\Phi_s \nonumber \\
    &\hspace{2cm} + \dfrac{3i}{8} \left( 2|\tilde{\psi}_s|^2 + \tilde{\rho}_{\as,s} + \dfrac{k^4}{3m^4a_s^4} \right)\delta\tilde{\psi}_s + \dfrac{3i}{16}\tilde{\psi}^2_s\delta\tilde{\psi}^*_s + \dfrac{ik^2}{4m^2a_s^2} \tilde{\psi}_s\Phi_s + \dfrac{1}{2}\tilde{\psi}_s\left( \Phi'_s + 3\Psi'_s \right), \\
    &\Psi'_s = -\tilde{H}_s\Phi_s + \dfrac{i}{4} \left(\tilde{\psi}^*_s\delta\tilde{\psi}_s - \tilde{\psi}_s\delta\tilde{\psi}^*_s \right) - \dfrac{1}{2}\delta\tilde{U}_{\as,s} - \dfrac{3}{8}\tilde{H}_s \left(\tilde{\psi}^*_s\delta\tilde{\psi}_s + \tilde{\psi}_s\delta\tilde{\psi}^*_s \right) - \dfrac{ik^2}{16m^2a_s^2} \left(\tilde{\psi}^*_s\delta\tilde{\psi}_s - \tilde{\psi}_s\delta\tilde{\psi}^*_s\right), \\
    &\dfrac{k^2}{m^2a_s^2}\Psi_s = - \dfrac{1}{2}\left(\tilde{\psi}^*_s\delta\tilde{\psi}_s + \tilde{\psi}_s\delta\tilde{\psi}^*_s + \delta\tilde{\rho}_{\as,s} \right) -\dfrac{3i}{4}\tilde{H}_s \left(\tilde{\psi}^*_s\delta\tilde{\psi}_s - \tilde{\psi}_s\delta\tilde{\psi}^*_s\right) + \dfrac{1}{2}|\tilde{\psi}_s|^2\Phi_s + \dfrac{3}{2}\tilde{H}_s\delta\tilde{U}_{\as,s}.
\end{align}
\end{slow_eqs}
\end{widetext}
Since the background fields are gauge-independent, the above equations should be supplemented with the same equations given in \eqref{eq:slow_psi}, \eqref{eq:slow_H} for completeness.

\section{Conclusion and outlook} \label{sec:conclusion}

This paper provides a comprehensive study for the effective field theory of the axion field in realistic cosmological context. The most important results we have obtained are dynamical equations and a fluid description for the slow modes in the synchronous and Newtonian gauge, as highlighted in Sec.~\ref{sec:complete_eft_syn} and Sec.~\ref{sec:complete_eft_New}.

At first glance, the EFT formalism may appear intimidating, as it requires two layers of expansion for every dynamical variable, as suggested by \eqref{eq:nu_expansion} and \eqref{eq:epsilon_expansion}. However, this complexity comes with several benefits. The $\nu$-expansion \eqref{eq:nu_expansion} ensures that any oscillations induced by the axion field are robustly captured and can be reconstructed, a feature that is particularly unique to this formalism. The $\epsilon$-expansion \eqref{eq:epsilon_expansion}, on the other hand, allows for precision control of the effective theory through relativistic corrections. This expansion also reduces otherwise intractable exact mode equations to a series of easier problems of finding corrections in terms of the slow modes. Both expansions make the EFT formalism highly systematic, such that its application to other axion theories can be carried out by simply following a similar procedure outlined in the main text. We believe that the present work has laid a solid foundation for such a task.

Regarding the fluid description for the axion field, we emphasize again that the slow-mode perturbations of the axion density, pressure and velocity derived in this work should not be confused with the effective fluid quantities of~\cite{Salehian:2020bon}. The former characterize the averaged properties of a perfect fluid describing the exact axion field, and are therefore always well-defined. The latter, by contrast, are identified from imperfect fluid equations with finite viscosity, and appear to be ill-defined when the axion field co-exists with other species. This distinction is, however, found to be insignificant unless high-order corrections are taken into account.

To conclude, let us remind that in the companion paper with this one~\cite{Luu:2026bzr}, we develop a completely new numerical package based on \texttt{CLASS} to derive axion perturbations and other cosmological observables using the EFT formalism. A correct numerical method should distinguish exact variables from their slow-mode counterparts as they follow different set of equations. We will demonstrate that, when implemented correctly, the EFT results yield an excellent reconstruction of the axion field’s exact oscillatory behavior, achieving up to subpercent agreement even with modest corrections. Moving forward, this direct application provides strong motivation for further investigation of the EFT formalism.

\begin{acknowledgments}
We are thankful to Mohammad Hossein Namjoo and Patrick Fitzpatrick for helpful comments and conversations. We also thank Luna Zagorac, Nathan Musoke, Mark Neyrinck, David Kaiser, Risa Wechsler, and Priya Natarajan, for encouraging conversations along the way. We thank the administrative and facilities staff at the University of New Hampshire including Katie Makem-Boucher and Michelle Mancini. HNL and CPW's contributions to this project were supported by DOE Grant DE-SC0025365, and we gratefully acknowledge the work of the DOE program officers who support the management of this grant.
\end{acknowledgments}

\appendix

\section{Slow modes of non-axion species} \label{app:non-axion_slow}

In the main text, we focus on the derivation of the slow-mode equations for the axion field and the metric perturbations. This appendix briefly discusses the equations governing non-axion species. As there are in principle infinitely many ways in which new physics can enter the system, we restrict our attention to a few concrete examples involving the most essential species in the $\Lambda$CDM model.

It is straightforward to see that any slow-mode equations not directly governing the evolution of the axion field variables take the same form as their exact counterparts, to the leading order. For example, from \eqref{eq:slow_eta} and \eqref{eq:slow_h}, the equations for $\eta_s$ and $h'_s$ can be written as
\begin{align}
    &\eta'_s = -\dfrac{1}{2}\delta \tilde{U}_{a,s} - \dfrac{1}{2}\delta\tilde{U}_{\as,s} + \mathcal{O}(\epsilon^3), \label{eq:slow_eta_lead} \\
    &\tilde{H}_sh'_s = \dfrac{2k^2}{m^2a_s^2}\eta_s + \delta\tilde{\rho}_{a,s} + \delta\tilde{\rho}_{\as,s} + \mathcal{O}(\epsilon^3), \label{eq:slow_h_lead}
\end{align}
which look identical to \eqref{eq:eta_exact} and \eqref{eq:hprime_exact}, with all exact variables replaced by their corresponding slow modes. This observation reflects the fact that the backreaction from axion oscillations should not significantly alter the dynamics of other variables. Thus, we expect that all equations governing non-axion species remain unchanged at the leading order. 

To further prove this point, let us expand the energy and momentum conservation equations from Eq.~\eqref{eq:non-axion_energy} and Eq.~\eqref{eq:non-axion_momentum} to obtain
\begin{align}
    &\delta\tilde{\rho}'_{\as,\nu} + i\nu\tilde{\rho}_{\as,\nu} = - 3\tilde{H}_\alpha \left( \delta\tilde{\rho}_{\as,\nu-\alpha} + \delta\tilde{p}_{\as,\nu-\alpha} \right) \nonumber \\ 
    &\hspace{1.7cm}  + \dfrac{k^2}{m^2}r_\alpha \delta\tilde{U}_{\as,\nu-\alpha} - \dfrac{1}{2}(\tilde{\rho}_{\as,\alpha} + \tilde{p}_{\as,\alpha}) \nonumber \\ 
    &\hspace{3.2cm} \times \left[ h'_{\nu-\alpha} + i(\nu-\alpha)h_{\nu -\alpha} \right], \\
    &\delta \tilde{U}'_{\as,\nu} + i\nu\tilde{U}_{\as,\nu} = - 3\tilde{H}_\alpha\delta\tilde{U}_{\as,\nu-\alpha} -\delta\tilde{p}_{\as,\nu} + \tilde{\sigma}_{\as,\nu}.
\end{align}
Setting $\nu = 0$ in the above equations, we have
\begin{align}
    &\delta\tilde{\rho}'_{\as,s} = - 3\tilde{H}_s \left( \delta\tilde{\rho}_{\as,s} + \delta\tilde{p}_{\as,s} \right) \nonumber \\ 
    &\hspace{1cm} + \dfrac{k^2}{m^2a_s^2} \delta\tilde{U}_{\as,s} - \dfrac{1}{2}(\tilde{\rho}_{\as,s} + \tilde{p}_{\as,s})h'_s + \mathcal{O}(\epsilon^6), \label{eq:non-axion_drho_s} \\
    &\delta \tilde{U}'_{\as,s} = - 3\tilde{H}_s\delta\tilde{U}_{\as,s} -\delta\tilde{p}_{\as,s} + \tilde{\sigma}_{\as,s} + \mathcal{O}(\epsilon^6). \label{eq:non-axion_dU_s}
\end{align}
These slow-mode equations are indeed consistent with \eqref{eq:non-axion_energy} and \eqref{eq:non-axion_momentum}, up to two orders beyond the leading one. In comparison to the metric equations \eqref{eq:slow_eta} and \eqref{eq:slow_h} where axion backreactions enter as next-to-leading order corrections, they appear to be even more strongly suppressed in the equations governing non-axion species. This is expected, as metric perturbations are directly sourced by axion perturbations, whereas non-axion perturbations couple to axion perturbations only through the metric perturbations.

Explicitly, let us consider the exact equations for the density perturbations of photons, baryons and CDM in the synchronous gauge, given by~\cite{Ma:1995ey}
\begin{align}
    &\dot{\delta}_\gamma = - \dfrac{4}{3a}\theta_\gamma - \dfrac{2}{3}\dot{h} \nonumber \\ 
    &\hspace{0.5cm} \rightarrow \Hquad \delta\tilde{\rho}'_\gamma = - 4\tilde{H}\delta\tilde{\rho}_\gamma + \dfrac{k^2}{a^2m^2} \delta\tilde{U}_\gamma - \dfrac{2}{3}\bar{\rho}_\gamma h', \\
    &\dot{\delta}_b = - \dfrac{1}{a} \theta_b - \dfrac{1}{2}\dot{h} \nonumber \\ 
    &\hspace{0.5cm} \rightarrow \Hquad \delta\tilde{\rho}'_b = - 3\tilde{H}\delta\tilde{\rho}_b + \dfrac{k^2}{a^2m^2} \delta\tilde{U}_b - \dfrac{1}{2}\bar{\rho}_b h', \\
    &\dot{\delta}_c  = -\dfrac{1}{2}\dot{h}_c \nonumber \\ 
    &\hspace{0.5cm} \rightarrow \Hquad \delta\tilde{\rho}'_c = - 3\tilde{H}\delta\tilde{\rho}_c -\dfrac{1}{2}\bar{\rho}_c h',
\end{align}
where the subscripts ``$\gamma$'', ``$b$'', ``$c$'' are used to denote fluid quantities of photons, baryons and CDM, respectively. These exact equations follow the same form as Eq.~\eqref{eq:non-axion_energy}, albeit with different EOS\footnote{The baryon EOS has been assumed to be negligible, {\it i.e.}, $w_b \sim 0$.}. Thus, our derivation leading to the generalized slow-mode equation \eqref{eq:non-axion_drho_s} for non-axion species applies equally here
\begin{align}
    &\delta\tilde{\rho}'_{\gamma,s} = - 4\tilde{H}_s\delta\tilde{\rho}_{\gamma,s} + \dfrac{k^2}{m^2a_s^2} \delta\tilde{U}_{\gamma,s} - \dfrac{2}{3}\bar{\rho}_{\gamma,s} h'_s + \mathcal{O}(\epsilon^6), \\
    &\delta\tilde{\rho}'_{b,s} = - 3\tilde{H}_s\delta\tilde{\rho}_{b,s} + \dfrac{k^2}{m^2a^2_s} \delta\tilde{U}_{b,s} - \dfrac{1}{2}\bar{\rho}_{b,s} h'_s + \mathcal{O}(\epsilon^6), \\
    &\delta\tilde{\rho}'_{c,s} = - 3\tilde{H}\delta\tilde{\rho}_{c,s} -\dfrac{1}{2}\bar{\rho}_{c,s} h'_s + \mathcal{O}(\epsilon^6).
\end{align}
While this result may seem trivial, it confirms that no modification to the non-axion equations is required when the axion field is included, as one simply needs to treat their slow-mode variables as the exact ones.

That said, the error estimation of order $\mathcal{O}(\epsilon^6)$ relies on the fact that $\delta \tilde{U}_{\as,\nu}^{(2)}$ and $\delta \tilde{U}_{\as,\nu}^{(3)}$ actually vanish, which in turn assumes $\tilde{\sigma}_{\as,\nu}^{(2)}$ and $\tilde{\sigma}_{\as,\nu}^{(3)}$ to be negligible in Eq.~\eqref{eq:delta_U_nu_2_3}. One may wonder whether this assumption can be justified. To address this concern, let us examine the exact equation governing the photon shear, which reads
\begin{align}
    2\dot{\sigma}_\gamma &= \dfrac{8}{15a}\theta_\gamma - \dfrac{3k}{a}F_{\gamma3} + \dfrac{4}{15}\dot{h} + \dfrac{8}{5}\dot{\eta} \nonumber \\ 
    &\hspace{1cm} - \dfrac{9}{5}n_e\sigma_T\sigma_\gamma + \dfrac{1}{10}n_e\sigma_T(G_{\gamma0} + G_{\gamma2}) \\
    \rightarrow  \tilde{\sigma}'_\gamma &= - \dfrac{4k^2}{15m^2a^2}\delta\tilde{U}_\gamma + \dfrac{8}{45}\tilde{\rho}_\gamma h' + \dfrac{16}{15}\tilde{\rho}_\gamma \eta' \nonumber \\
    & - \dfrac{9}{10}\tilde{n}_e\tilde{\sigma}_\gamma - \dfrac{3k}{2ma}\tilde{F}_{\gamma3} + \dfrac{1}{20}\tilde{n}_e(\tilde{G}_{\gamma0} + \tilde{G}_{\gamma2}), \label{eq:sigma_gamma}
\end{align}
where $n_e$ is the time-dependent electron number density, $F_{\gamma l}$ and $G_{\gamma l}$ denote high-$l$ multipoles of the photon phase-space density distribution (see \cite{Ma:1995ey} for their exact definition). For convenience, we have also introduced the following notations
\begin{align}
\begin{gathered}
    \tilde{n}_e = \dfrac{n_e\sigma_T}{m}, \quad \tilde{F}_{\gamma l} = (\tilde{\rho}_\gamma + \tilde{p}_\gamma)F_{\gamma l}, \\ \tilde{G}_{\gamma l} = (\tilde{\rho}_\gamma + \tilde{p}_\gamma)G_{\gamma l}.
\end{gathered}
\end{align}
Note that Eq.~\eqref{eq:sigma_gamma} usually applies when baryons and photons have decoupled after recombination. At this time, the Thomson scattering term becomes comparable to the Hubble rate, {\it i.e.}, $n_e\sigma_T \sim H$, so it is reasonable to estimate $\tilde{n}_e \sim \tilde{H} \sim \mathcal{O}(\epsilon)$\footnote{In contrast, the estimation must be carried out with a different set of equations before recombination, where $\tilde{n}_e \gg \tilde{H}$, and the tight-coupling approximation is typically invoked to treat photons and baryons as a single fluid, which we do not pursue further here.}. As for the photon multipoles, we can assume $\tilde{F}_{\gamma l} \sim \tilde{G}_{\gamma l} \sim \mathcal{O}(\epsilon^2)$ at least, since $\tilde{F}_{\gamma 2} \equiv 2\tilde{\sigma}_\gamma \sim \mathcal{O}(\epsilon^2)$. 

That said, in practice, it is not necessary to know the precise magnitudes or equations governing these variables in order to estimate $\tilde{\sigma}_{\gamma,\nu}^{(2)}$ and $\tilde{\sigma}_{\gamma,\nu}^{(3)}$. With some physical intuition, we know that the lowest-order corrections to non-axion variables only arise from sources with axion oscillations. For instance, $h^{(2)}_\nu$ as the second-order correction to $h_\nu$ is non-vanishing, as shown in \eqref{eq:h_nu_2}, because its exact equation \eqref{eq:h_dot_dot} is directly driven by the axion pressure $\delta\tilde{p}_a$. With a simple reconstruction using \eqref{eq:delta_p_nu_2_3}, \eqref{eq:slow_delta_p_a} and \eqref{eq:nu_expansion}, it is clear that this variable is oscillatory even at leading order
\begin{align}
    \delta\tilde{p}_a = - \tilde{\psi}_s\delta\tilde{\psi}_se^{-2imt} - \tilde{\psi}^*_s\delta\tilde{\psi}^*_se^{2imt} + \mathcal{O}(\epsilon^3).
\end{align}

Following the same logic, we can argue with high confidence that any significant corrections to $\tilde{\sigma}_{\gamma,\nu}$ (at the lowest orders) must originate from $h'$ and $\eta'$ terms in Eq.~\eqref{eq:sigma_gamma}. In other words, if these two terms were removed from \eqref{eq:sigma_gamma}, the corrections to $\tilde{\sigma}_\nu$ would vanish at every order, since axions cannot induce backreaction on photons in the absence of any coupling between them\footnote{Mathematically speaking, couplings through $\tilde{H}$, which appears in the recombination equation for $\tilde{n}_e$, may still exist, but these are expected to contribute only at very high orders.}. In this context, expanding Eq.~\eqref{eq:sigma_gamma} while neglecting the last three terms yields
\begin{multline}
    \tilde{\sigma}'_{\gamma,\nu} + i\nu\tilde{\sigma}_{\gamma,\nu} = - \dfrac{4k^2}{15m^2}r_\alpha \delta\tilde{U}_{\gamma,\nu-\alpha} \nonumber \\ + \dfrac{8}{45}\tilde{\rho}_{\gamma,\alpha} \left[ h'_{\nu-\alpha} + i(\nu-\alpha)h_{\nu-\alpha} \right] \\ + \dfrac{16}{15}\tilde{\rho}_{\gamma,\alpha} \left[ \eta'_{\nu-\alpha} + i(\nu-\alpha)\eta_{\nu-\alpha} \right].
\end{multline}
With $\nu \neq 0$, the second- and third-order corrections to $\tilde{\sigma}_{\gamma,\nu}$ can now be derived as
\begin{align}
    &i\nu\tilde{\sigma}^{(2)}_{\gamma,\nu} = \dfrac{8i\nu}{45}\tilde{\rho}_{\gamma,s} h^{(0)}_\nu, \\
    &\tilde{\sigma}^{(2)'}_{\gamma,\nu} +i\nu\tilde{\sigma}^{(3)}_{\gamma,\nu} = -\dfrac{4k^2}{16m^2a_s^2}\delta\tilde{U}^{(2)}_{\gamma,\nu} \nonumber \\
    &\hspace{2cm} + \dfrac{8i\nu}{45}\tilde{\rho}_{\gamma,s} h^{(1)}_\nu + \dfrac{16i\nu}{15}\tilde{\rho}_{\gamma,s} \eta^{(1)}_\nu.
\end{align}
As $h^{(0)}_\nu = 0$, the first equation gives $\tilde{\sigma}^{(2)}_{\gamma,\nu} = 0$, which in turn implies $\delta U^{(2)}_{\gamma,\nu} = 0$ from \eqref{eq:delta_U_nu_2_3}. Substituting these results into the second equation, along with $h^{(1)}_\nu = 0$ and $\eta^{(0)}_\nu = 0$, we find $\tilde{\sigma}^{(3)}_{\gamma,\nu} = 0$, which similarly implies $\delta U^{(3)}_{\gamma,\nu} = 0$ from \eqref{eq:delta_U_nu_2_3}. Although we have used photons as an example here, the same results are expected to apply to the shear of relativistic neutrinos as well.


\bibliography{reference}

\end{document}